\documentclass{article}
\usepackage[utf8]{inputenc}
\usepackage{authblk}

\title{Triple-Spherical Bessel Function Integrals with Exponential and Gaussian Damping: Towards An Analytic N-Point Correlation Function Covariance Model}

\author[1*]{Jessica Chellino}
\author[1,2]{Zachary Slepian}
\affil[1]{\footnotesize Department of Astronomy, University of Florida, 211 Bryant Space Science Center, Gainesville, FL 32611, USA}
\affil[2]{\footnotesize Physics Division, Lawrence Berkeley National Laboratory, 1 Cyclotron Road, Berkeley, CA 94709, USA}
\affil[*]{Electronic Address: {jchellino@ufl.edu}}

\usepackage[a4paper, total={6in, 8in}]{geometry}
\usepackage{amsmath}
\usepackage{amssymb}

\usepackage[dvipsnames]{xcolor}

\usepackage[numbers]{natbib}
\usepackage{graphicx}
\usepackage{subcaption}
\usepackage{color}
\usepackage{enumitem}
\def\beq{\begin{eqnarray}}
\def\eeq{\end{eqnarray}}

\definecolor{newgreen}{RGB}{0,200,0}
\definecolor{purple}{RGB}{90,0,230}

\usepackage{etoolbox}% http://ctan.org/pkg/etoolbox
\makeatletter
\patchcmd{\math@cr@@@align}% <cmd>
  {\place@tag}% <search>
  {\bgroup\color{black}\place@tag\egroup}% <replace>
  {}{}% <success><failure>
\makeatother

\begin{document}

\maketitle

\begin{abstract}
Spherical Bessel functions (sBFs) appear commonly in many areas of physics wherein there is both translation and rotation invariance, and often integrals over products of several arise. Thus, analytic evaluation of such integrals with different weighting functions (which appear as toy models of a given physical observable, such as the galaxy power spectrum) is useful. Here, we present a generalization of a recursion-based method for evaluating such integrals. It gives relatively simple closed-form results in terms of Legendre functions (for the exponentially-damped case) and Gamma, incomplete Gamma, and hypergeometric functions (for the Gaussian-damped case). We also present a new, non-recursive method to evaluate integrals of products of sBFs with Gaussian damping in terms of incomplete Gamma functions and hypergeometric functions.
\end{abstract}

\section{Introduction}
\label{sec:intro}
Spherical Bessel functions (sBFs) are ubiquitous in cosmology. The three-dimensional plane wave offers an eigenbasis for the gradient operators that appear in \textit{e.g.} cosmological perturbation theory \cite{bern_pt}, and in general the plane waves are the eigenstates of momentum, which is the conserved quantity associated by Noether's theorem with the translation symmetry implied by the cosmological assumption of statistical homogeneity of the large-scale structure (LSS). However, cosmology also assumes isotropy (rotation symmetry) about any point, so spherical coordinates are often apt. They do not diagonalize the fluid equations as well (though efforts to solve these in the spherical basis have been made; \textit{e.g.} \cite{hoffman}), but they are especially useful for numerical computation both of CMB anisotropies \cite{seljak} and lensing \cite{assassi} and LSS perturbative corrections \cite{schmitt_1, schmitt_2, rotation, pt_decoup, philcox_to_sum}. Efficient algorithms exist for performing sBF transforms numerically \cite{hamilton, mcewen, grass_jeong, Fang, tom_jeong, grass_dore, khek}. sBFs have also been explored as the ``correct'' basis for wide-angle redshift-space distortion (RSD) work \cite{culhane} and, in a modified form, for LSS surveys with finite redshift width (leading to a spherical shell survey geometry) \cite{lado_sbf}. 

sBFs also commonly appear in other areas of physics. Mie scattering, in which electromagnetic waves are scattered off of a sphere, has scattering coefficients in terms of Riccati-Bessel functions \cite{qjamm_mie, light_scatt, scatt_sphere}. It is necessary to evaluate integrals of the product of sBFs in many scattering calculations (\textit{e.g.} \cite{atomic_num, int_comp_scatt, bremsstrahlung, non_rel_scatt, new_sbf_quantum}). Such integrals may include Gaussian damping to ensure convergence at infinity. Modified Bessel functions often arise in the Ewald summation, which allows long-range interactions to be computed \cite{ewald_yukawa, ewald_sum}. Additionally, sBF integrals are relevant to spherical linear canonical transforms, which can be applied to studies of acoustics, optics and heat \cite{sph_lin_canon}.

The plane wave expansion links the plane wave to the sBFs and spherical harmonics, and provides a powerful ``Rosetta stone'' for translating between situations where translation invariance is the more important symmetry and situations where rotation invariance is more germane. 

While efficient numerical algorithms exist for transforms of one, two, and even three \cite{rotation} sBFs against a source function given as a numerical lookup table, there are occasions where the integrand is given in closed form and thus an analytic evaluation of sBF integrals is of interest. For instance, in computations of the covariance matrix for pair correlations (2-Point Correlation Function, 2PCF), triplet correlations (3PCF, \textit{e.g.} \cite{SE_3PCF_BAO, sarabande}), 4PCF (\textit{e.g.} \cite{connected, cahn_parity, hou_parity}), or beyond, a power law model of the galaxy power spectrum often provides insight into the gross properties of these large matrices. Such models for the 2PCF were explored in \cite{cohn}, for the 3PCF briefly in \cite{3pt_alg}, \S6.3, and at more length in \cite{jiamin_covar}, appendix D (the general integrals used for covariance of 3PCF, 4PCF and beyond). At the signal level, triple-sBF integrals were used prominently in the computation of the redshift-space galaxy power spectrum of \cite{fonseca}, see in particular \S3.2 and appendix B. This work used a method of parametric differentiation developed by \cite{fabrikant}, based on Rayleigh's formula for the sBFs.

There are numerous works on the integral of a known function against a single-sBF (a recent, incomplete list with fuller referencing within is \cite{adkins_ft, indef_bloom}) and on the double-sBF case (more fully reviewed in \cite{meigs}), but far fewer regarding triple-sBF integrals. Early work is in Watson's \textit{Treatise}, \cite{wat}, with further results in terms of Appell F4 functions in \cite{bailey} (which simplify in some cases) and in \cite{jackson1972integrals} (reduction to products of associated Legendre polynomials of the external angles of a triangle formed by the free arguments) and \cite{mehrem_42, mehrem_3_gen}, in terms of Legendre polynomials in the cosine of the internal enclosed angle of that triangle. These results are for a somewhat general power law (\cite{bailey}, leading to the Appell F4, and simplifying if the power is a particular combination of the sBF orders) or a $k^2$ power law, \textit{i.e.} the usual spherical coordinates Jacobian. Gervois and Navelet \cite{gervois89} pursues an integer power law and is subject to some restrictions on the sBF orders. There have been recent works in cosmology pursuing a variety of methods: \cite{chen04, shellard}, and most relevant for the current work, \cite{wk}. This last work developed a recursion for triple-sBF integrals with a square power law weight, using the three-function recursion relation for the sBFs. 

In the current work, we extend this recursion method to the case of a triple-sBF integral with an exponential damping and then with a Gaussian damping. Damped exponentials and Gaussians are both useful toy models for the galaxy power spectrum, and have the property that they lead to simple, finite 2PCFs, and that the integrals associated with the analytic covariance matrix formalism of \cite{jiamin_covar} are all convergent. In a second paper, we explore an analytic covariance matrix model for the 2PCF-4PCF using these toy power spectra more fully. In the course of that work we found it necessary to develop methods to evaluate these integrals analytically, which we present here.

\section{Triple-sBF Integrals with Exponential Damping}
\label{sec:exp}

\subsection{Recursion Relation Approach}
\label{sssec:exp_rec}
A recursion relation has been used to evaluate triple-sBF integrals \cite{wk}. Here we extend this recursion to evaluate such integrals including an exponential damping term. We will find that the result involves Legendre functions of the second kind.

We wish to evaluate integrals of the form 
\begin{align}
\label{eqn:exp_general}
   f_{\ell_1,\ell_2,\ell_3}(r_1,r_2,r_3;p) = \int_{0}^{\infty}k^2dk\;e^{-p^2k} j_{\ell_1}(kr_1)j_{\ell_2}(kr_2)j_{\ell_3}(kr_3)
\end{align} for $\{p, \ell_i, r_i\} \in \mathbb{R}$, $\ell_i \in \mathbb{Z}$, $\ell_i \geq 0$, $r_1 > 0$, $r_i \neq 0$, and $p \neq 0$, as will be explained below. We begin with the recursion relation \citep{NIST:DLMF} for spherical Bessel functions:
\begin{align}
\label{eqn:sbf_recursion}
j_{\ell-1}(x) + j_{\ell+1}(x) = \frac{2\ell+1}{x}j_\ell(x),
\end{align}
where $\ell$ must be a positive integer. This recursion can be extended \cite{wk} to a recursion for triple-sBF integrals as
{\color{newgreen}
\begin{align}
\label{eqn:int_recursion}
f_{\ell_1,\ell_2+1,\ell_3} = \left(\frac{r_1}{r_2}
\right)\left(\frac{2\ell_2+1}{2\ell_1+1}\right)\left(f_{\ell_1-1,\ell_2,\ell_3} + f_{\ell_1+1,\ell_2,\ell_3}\right) - f_{\ell_1,\ell_2-1,\ell_3}\textcolor{black}{.}
\end{align}} 
To use this recursion to evaluate all cases, we require three base cases to anchor it: $f_{\ell,0,0}$, $f_{\ell,-1,0}$, and $f_{\ell,-1,-1}$. The first base case is 
\begin{align}
f_{\ell,0,0}(r_1,r_2,r_3;p) = \int_{0}^{\infty}k^2dk\;e^{-p^2k}j_{\ell}(kr_1)j_{0}(kr_2)j_{0}(kr_3),
\end{align}
from setting $\ell_2 = 0 = \ell_3$ in equation (\ref{eqn:exp_general}). By representing the first spherical Bessel function in the integrand  (order $\ell$, argument $k r_1$) as in \cite{mehrem_42}, we obtain
\begin{align}
\label{eqn:sbf_as_Pl}
j_{\ell}(kr_1) = \frac{(-i)^\ell}{2}\int_{-1}^{1}d\mu\;P_\ell(\mu)e^{ikr_1\mu},
\end{align}
which requires $r_1 \geq 0$. The first base case becomes
\begin{align}
\label{eqn:exp_with_sines}
    f_{\ell,0,0}(r_1,r_2,r_3;p) = \frac{(-i)^\ell}{2r_2r_3}\int_{-1}^{1}d\mu\;P_\ell(\mu)\int_{0}^{\infty}dk\;e^{ikr_1\mu-p^2k}\sin(kr_2)\sin(kr_3).
\end{align}
Each $j_0$ in equation (\ref{eqn:sbf_as_Pl}) has been rewritten explicitly (with $k$ in the denominator of each) to cancel the $k^2$ in the Jacobian; we require $r_2, r_3 \neq 0$.

After rewriting the sines in equation (\ref{eqn:exp_with_sines}) as complex exponentials, we find
\begin{align}    
\label{eqn:seven}
    f_{\ell,0,0}(r_1,r_2,r_3;p) &= -\frac{1}{4r_2r_3}\frac{(-i)^\ell}{2}\int_{-1}^{1}d\mu\;P_\ell(\mu)\int_{0}^{\infty}dk\;e^{-p^2k}\left[e^{-ik(-r_1\mu-r_2-r_3)}-e^{-ik(-r_1\mu-r_2+r_3)} \nonumber \right.  \\
    & \left. \qquad\qquad\qquad\qquad\qquad\qquad\qquad\qquad  -e^{-ik(-r_1\mu+r_2-r_3)}+e^{-ik(-r_1\mu+r_2+r_3)}\right].
\end{align}
As long as $p \neq 0$, we can use \citep{GR} equation 3.310 to evaluate the $k$ integral, obtaining
\begin{align} 
    f_{\ell,0,0}(r_1,r_2,r_3;p) &= -\frac{1}{4r_2r_3}\frac{(-i)^\ell}{2}\int_{-1}^{1}d\mu\;P_\ell(\mu)\left[\left[p^2+i\left(-r_1\mu-r_2-r_3\right)\right]^{-1} \nonumber \right.\\
    &\left. \qquad -\left[p^2+i\left(-r_1\mu-r_2+r_3\right)\right]^{-1} -\left[p^2+i\left(-r_1\mu+r_2-r_3\right)\right]^{-1} \nonumber \right.\\
    &\left. \qquad +\left[p^2+i\left(-r_1\mu+r_2+r_3\right)\right]^{-1}\right]. 
\end{align}
We then use \cite{GR} equation 7.224.1 to integrate over $\mu$, resulting in
{\color{blue}
\begin{align}
\label{eqn:exp_bc1}
    f_{\ell,0,0}(r_1,r_2,r_3;p) &= -\frac{1}{4r_1r_2r_3}(-i)^{\ell+1}\Big[Q_\ell(R_{--})-Q_\ell(R_{-+})-Q_\ell(R_{+-})+Q_\ell(R_{++})\Big] 
\end{align}}
where $r_1 \neq 0$, $Q_{\ell}$ is the order-$\ell$ Legendre function of the second kind and $R_{\pm\pm}$ is defined as
\begin{align}
\label{eqn:R_def}
    R_{\pm\pm} &\equiv \frac{1}{r_1}\left(-i p^2 \pm r_2 \pm r_3\right).
\end{align}
We note that \citep{GR} equation 7.224.1 has a discontinuity when the argument of the $Q_{\ell}$ is along the real line from $-1$ to $+1$, but the arguments $R_{\pm \pm}$ we have of the $Q_{\ell}$ above are manifestly complex ($\{p, r_i\} \in \mathbb{R}$) and hence do not lie along this line.

The other base cases can be obtained in the same way, resulting in 
{\color{blue}
\begin{align}
\label{eqn:exp_bc2}
f_{\ell,-1,0}(r_1,r_2,r_3;p) &= -\frac{1}{4r_1r_2r_3}(-i)^{\ell} \Big[Q_\ell(R_{--}) - Q_{\ell}(R_{-+})+Q_{\ell}(R_{+-}) - Q_{\ell}(R_{++})\Big]
\end{align}} 
and
{\color{blue}
\begin{align}
\label{eqn:exp_bc3}
f_{\ell,-1,-1}(r_1,r_2,r_3;p) &= \frac{1}{4r_1r_2r_3}(-i)^{\ell+1}\Big[Q_{\ell}(R_{--}) + Q_{\ell}(R_{-+})+Q_{\ell}(R_{+-}) + Q_{\ell}(R_{++})\Big]\textcolor{black}{.}
\end{align}}

The recursion relation (\ref{eqn:int_recursion}) may then be used to construct from these base cases the integral (\ref{eqn:exp_general}) for any other non-negative integer values of the $\ell_i$. We note that equations (\ref{eqn:exp_bc1}), (\ref{eqn:exp_bc2}), and (\ref{eqn:exp_bc3}) depend on the same four Legendre functions, although the signs in front of each term differ. The sign differences as well as the different complex pre-factors arise from writing sines and cosines as complex exponentials: equation (\ref{eqn:exp_bc1}) depends on the product of two sines, equation (\ref{eqn:exp_bc2}) the product of a sine and cosine, and equation (\ref{eqn:exp_bc3}) the product of two cosines.

\subsection{Proof That the Base Cases Are Real}
\label{sssec:q_ell}

We now show that the results obtained for the three base cases are real; they should be real since the beginning integrands and domain of integration for each of them are real. We note that for each base case (equations \ref{eqn:exp_bc1}, \ref{eqn:exp_bc2}, and \ref{eqn:exp_bc3}), the argument of the fourth $Q_{\ell}$ is negative the complex conjugate of that of the first, and similarly for the third and the second. For $f_{\ell,0,0}$ and $f_{\ell,-1,-1}$ we therefore must show that $(-i)^{\ell+1}\left[Q_\ell(-z^*)+Q_\ell(z)\right]$ is real (this identity applies to each pair of $Q_{\ell}$ as above). For $f_{\ell,-1,0}$, we need to demonstrate that $(-i)^{\ell}\left[Q_\ell(-z^*)-Q_\ell(z)\right]$ is real (and again, this identity then applies to each pair of $Q_{\ell}$ noted above). 

We write $z = re^{i\theta}$ and expand the Legendre function as
\begin{align}
    Q_\ell(z) = \sum\limits_{\text{odd\;} n}c_nz^{-(\ell+n)},
\end{align}
\textit{i.e.} as an infinite series. The more familiar representation is as a finite polynomial in $z$ multiplied by $\ln[(1 + z)/(1-z)]$, and it is this singular second factor that when expanded renders the power series infinite.

For $f_{\ell,0,0}$ and $f_{\ell,-1,-1}$ we then have 
\begin{align}
    (-i)^{\ell+1}\Big[Q_\ell(-z^*)+Q_\ell(z)\Big] &= (-i)^{\ell+1}\sum\limits_{\text{odd\;}n}c_nr^{-(\ell+n)}\left[\left(-e^{-i\theta} \right)^{-(\ell+n)} + \left(e^{i\theta} \right)^{-(\ell+n)} \right],
\end{align}
which for odd $\ell$, by taking the even part of the exponential  becomes
 \begin{align}   
    (-i)^{\ell+1}\Big[Q_\ell(-z^*)+Q_\ell(z)\Big] &= -2\sum\limits_{\text{odd n}}c_nr^{-(\ell+n)}\text{cos}\left[\theta \left(\ell+n \right)\right] \qquad\qquad \text{(odd}\; \ell),
\end{align}
and for even $\ell$, by taking the odd part of the exponential becomes
\begin{align}
    (-i)^{\ell+1}\Big[Q_\ell(-z^*)+Q_\ell(z)\Big]     &= 2\sum\limits_{\text{odd n}}c_nr^{-(\ell+n)}\text{sin}\left[\theta \left(\ell+n \right)\right] \qquad\qquad\;\;\;\; \text{(even}\;\ell).
\end{align}
Similarly, for $f_{\ell,-1,0}$ we have
\begin{align}
    (-i)^{\ell}\Big[Q_\ell(-z^*)-Q_\ell(z)\Big] &= \sum\limits_{\text{odd n}}c_nr^{-(\ell+n)}\left[\left(-e^{-i\theta} \right)^{-(\ell+n)} - \left(e^{i\theta} \right)^{-(\ell+n)} \right],
\end{align}
which by taking the even part of the exponential (for odd $\ell$) becomes
\begin{align}
    (-i)^{\ell}\Big[Q_\ell(-z^*)-Q_\ell(z)\Big] &= -2\sum\limits_{\text{odd n}}c_nr^{-(\ell+n)}\text{sin}\left[\theta \left(\ell+n \right)\right] \qquad\qquad\text{(odd}\; \ell),
\end{align}
and from the odd part of the exponential (for even $\ell$) is 
\begin{align}
    (-i)^{\ell}\Big[Q_\ell(-z^*)-Q_\ell(z)\Big] &= 2\sum\limits_{\text{odd n}}c_nr^{-(\ell+n)}\text{cos}\left[\theta \left(\ell+n \right)\right] \qquad\qquad\;\;\;\; \text{(even}\; \ell).
\end{align}
These results are manifestly real, which shows that all three base cases in \S\ref{sssec:exp_rec} must be real, as expected.

\section{Recursion Relation Approach for Triple-sBF Integrals with Gaussian Damping}
\label{sec:gauss_rec}

The integral of three sBFs with Gaussian damping can be evaluated by a method similar to that described in \S \ref{sssec:exp_rec}. To evaluate integrals of the form
\begin{align}
\label{eqn:gauss_rec}
   f_{\ell_1,\ell_2,\ell_3}(r_1,r_2,r_3;p) = \int_{0}^{\infty}k^2dk\;e^{-(pk)^2}j_{\ell_1}(kr_1)j_{\ell_2}(kr_2)j_{\ell_3}(kr_3)
\end{align} for $\{p, \ell_i, r_i\} \in \mathbb{R}$, $\ell_i \in \mathbb{Z}$, $\ell_i \geq 0$, $r_1 > 0$, $r_i \neq 0$, and $p \neq 0$, the same recursion relation given in equation (\ref{eqn:int_recursion}) \cite{wk} can be used with base cases $f_{\ell,0,0}$, $f_{\ell,-1,0}$, and $f_{\ell,-1,-1}$. We will find that the base cases can be expressed using regularized hypergeometric and incomplete Gamma functions. 

The first base case is from setting $\ell_2 = 0 =\ell_3$ in equation (\ref{eqn:gauss_rec}):
\begin{align}
f_{\ell,0,0}(r_1,r_2,r_3;p) = \int_{0}^{\infty}k^2dk\;e^{-(pk)^2}j_{\ell}(kr_1)j_{0}(kr_2)j_{0}(kr_3).
\end{align}
By using equation (\ref{eqn:sbf_as_Pl}) to rewrite the $j_{\ell}(kr_1)$, and writing the two  factors of $j_0$ explicitly to cancel the $k^2$ in the Jacobian, we obtain
\begin{align}
    f_{\ell,0,0}(r_1,r_2,r_3;p) &= \frac{(-i)^{\ell}}{2r_2r_3}\int_{-1}^1d\mu\;P_{\ell}(\mu)\int_0^{\infty}dk\;e^{-(pk)^2+ikr_1\mu} \sin(kr_2) \sin(kr_3),
\end{align}
where $\{r_2,r_3\} \neq 0$. After rewriting the sines as complex exponentials (obtaining the same set of exponentials as are in the square brackets in the integrand of equation \ref{eqn:seven}) and completing the square on what results, this becomes
\begin{align}
\label{eqn:comp_sq}
f_{\ell,0,0}(r_1,r_2,r_3;p) &= -\frac{(-i)^\ell}{8r_2r_3}\int_{-1}^1d\mu\;P_{\ell}(\mu) \nonumber \\
& \qquad \times \left\{\exp\left[-\left(\frac{R_{\mu ++}}{2p}\right)^2\right]\int_0^\infty dk\;\exp\left[-p^2\left(k-\frac{iR_{\mu ++}}{2p^2}\right)^2\right] \right. \nonumber \\
& \qquad -\exp\left[-\left(\frac{R_{\mu +-}}{2p}\right)^2\right]\int_0^\infty dk\;\exp\left[-p^2\left(k-\frac{iR_{\mu +-}}{2p^2}\right)^2\right] \nonumber \\
& \qquad -\exp\left[-\left(\frac{R_{\mu -+}}{2p}\right)^2\right]\int_0^\infty dk\;\exp\left[-p^2\left(k-\frac{iR_{\mu -+}}{2p^2}\right)^2\right]\nonumber \\
& \left. \qquad +\exp\left[-\left(\frac{R_{\mu --}}{2p}\right)^2\right]\int_0^\infty dk\;\exp\left[-p^2\left(k-\frac{iR_{\mu --}}{2p^2}\right)^2\right]\right\},
\end{align}
where we define 
\begin{align}
    R_{\mu \pm \pm} \equiv r_1\mu \pm r_2 \pm r_3.
\end{align}
We note that the ordering of signs on $r_2$ and $r_3$ in the integrand matches that outside the integral in each line because the factor outside is what needs to be subtracted off after completing the square in each integrand. 

The inner $k$ integrals can be evaluated using a change of variable $x = k-\left[(iR_{\mu \pm \pm})/(2p^2)\right]$ and \cite{GR} equation 2.33.16, as long as $p \neq 0$:
\begin{align}    
\int_{0}^{\infty}dk\;\exp\left[-p^2\left(k-\frac{iR_{\mu \pm \pm}}{2p^2}\right)^2\right] &= \int_{-(iR_{\mu \pm \pm})/(2p^2)}^{\infty}dx\;e^{-(px)^2} \nonumber \\
    &= \frac{\sqrt{\pi}}{2p}\left[1-\text{erf}\left(-\frac{iR_{\mu \pm \pm}}{2p} \right) \right].
\end{align}

We now must integrate over $\mu$ in equation (\ref{eqn:comp_sq}), where in the first line we have a Legendre polynomial in $\mu$. We expand this Legendre polynomial as\footnote{https://mathworld.wolfram.com/LegendrePolynomial.html, equation 32.}
\begin{align}
\label{eqn:leg_pow}
    P_{\ell}(\mu) = \frac{1}{2^{\ell}}\sum\limits_{n=0}^{\lfloor \ell/2 \rfloor}(-1)^n\binom{\ell}{n}\binom{2(\ell-n)}{\ell}\mu^{\ell-2n};
\end{align} 
this is simply a falling power series composed of powers of $\mu$ ranging from $\ell$ down to either one (if $\ell$ is odd) or zero (if $\ell$ is even) in even steps. We note that $\lfloor \ell/2 \rfloor$ is the floor function, which gives the largest integer less than or equal to its argument; if $\ell$ is even, $\ell/2$, but if $\ell$ is odd, $(\ell - 1)/2$.

The first base case then becomes
\begin{align} 
\label{eqn:26}
    f_{\ell,0,0}(r_1,r_2,r_3;p) &= -\frac{(-i)^{\ell}\sqrt{\pi}}{16pr_2r_3}\frac{1}{2^{\ell}}\sum\limits_{n=0}^{\lfloor \ell/2 \rfloor}(-1)^n\binom{\ell}{n}\binom{2(\ell-n)}{\ell}\int_{-1}^{1}d\mu\;\mu^{\ell-2n} \nonumber \\
    & \qquad \times\left\{\exp\left[-\left(\frac{R_{\mu ++}}{2p}\right)^2\right]\left[1-\text{erf}\left(-\frac{iR_{\mu ++}}{2p}\right) \right] \nonumber \right. \nonumber \\
    &\left. \qquad -\exp\left[-\left(\frac{R_{\mu +-}}{2p}\right)^2\right]\left[1-\text{erf}\left(-\frac{iR_{\mu +-}}{2p} \right) \right] \right. \nonumber \\
    &\left. \qquad -\exp\left[-\left(\frac{R_{\mu -+}}{2p}\right)^2\right]\left[1-\text{erf}\left(-\frac{iR_{\mu -+}}{2p} \right) \right] \right. \nonumber \\
    &\left. \qquad +\exp\left[-\left(\frac{R_{\mu --}}{2p}\right)^2\right]\left[1-\text{erf}\left(-\frac{iR_{\mu --}}{2p} \right) \right] \right\}.
\end{align}

The $\mu$ integrals in equation (\ref{eqn:26}) have the form 
\begin{align}
    I_{\mu}(r_1,r_2,r_3;p) \equiv \int_{-1}^{1}d\mu\;\mu^{\ell-2n}\exp\left[-\left(\frac{R_{\mu \pm \pm}}{2p}\right)^2\right]\left[1-\text{erf}\left(-\frac{iR_{\mu \pm \pm}}{2p} \right) \right].
\end{align}
After changing the integration variable from $\mu$ to $R_{\mu \pm \pm}$, then using the binomial expansion on $\mu^{\ell-2n}$, we find for $r_1 \neq 0$:
\begin{align}
\label{eqn:mu}
    I_{\mu}(r_1,r_2,r_3;p) &= r_1^{-1-\ell+2n}\sum\limits_{m=0}^{\ell-2n}\binom{\ell-2n}{m}\left(\mp r_2 \mp r_3 \right)^{\ell-2n-m} \nonumber \\
    & \qquad \times\int_{r_{- \pm \pm}}^{r_{+ \pm \pm}}dR_{\mu \pm \pm}\;R_{\mu \pm \pm}^me^{-[R_{\mu \pm \pm}/(2p)]^2}\left[1-\text{erf}\left(-\frac{iR_{\mu \pm \pm}}{2p} \right) \right].
\end{align}

Multiplying the two terms in the square bracket by the Gaussian damping in the integrand of equation (\ref{eqn:mu}) and then, in what results, focusing on the term proportional to unity, we have
\begin{align}
    \int_{r_{- \pm \pm}}^{r_{+ \pm \pm}}dR_{\mu \pm \pm}\;R_{\mu \pm \pm}^me^{-[R_{\mu \pm \pm}/(2p)]^2} &= -2^mp^{m+1}\left[\Gamma\left(\frac{m+1}{2},\left(\frac{r_{+ \pm \pm}}{2p}\right)^2\right) 
    \right. \nonumber \\
     & \left. \qquad\qquad\qquad\qquad\qquad\qquad\qquad -\Gamma\left(\frac{m+1}{2},\left(\frac{r_{- \pm \pm}}{2p}\right)^2\right)\right] \nonumber \\
    &= G_{- \pm \pm}^{(m)}(r_1,r_2,r_3;p) - G_{+ \pm \pm}^{(m)}(r_1,r_2,r_3;p)
\end{align}
from \cite{GR} equation 2.33.10 where $\Gamma$ is the incomplete Gamma function and we have defined
\begin{align}
\label{eqn:r_def}
    r_{\pm \pm \pm} &\equiv \pm r_1 \pm r_2 \pm r_3 
\end{align} 
and
\begin{align}
\label{eqn:G}
    G_{\pm \pm \pm}^{(m)}(r_1,r_2,r_3;p) &\equiv 2^mp^{m+1}\Gamma\left(\frac{m+1}{2},\left(\frac{r_{\pm \pm \pm}}{2p}\right)^2 \right).
\end{align}

Multiplying the two terms in the square bracket of equation (\ref{eqn:mu}) by the Gaussian damping in the integrand and then focusing on the term proportional to the error function, we obtain\footnote{https://functions.wolfram.com/GammaBetaErf/Erf/21/01/02/04/01/0007/}
\begin{align}
    &\int_{r_{- \pm \pm}}^{r_{+ \pm \pm}}dR_{\mu \pm \pm}\;R_{\mu \pm \pm}^m e^{-[R_{\mu \pm \pm}/(2p)]^2}\text{erf}\left(-\frac{iR_{\mu \pm \pm}}{2p} \right) = -\frac{i}{4p}\Gamma\left(\frac{m+2}{2}\right) \nonumber \\
    &\qquad\times\Bigg[r_{+ \pm \pm}^{m+2}\;_2{\tilde{F}_2}\left(1,\frac{m+2}{2};\frac{3}{2},\frac{m+4}{2};-\left(\frac{r_{+ \pm \pm}}{2p}\right)^2  \right) \Bigg. \nonumber \\ 
    &\Bigg. \qquad - r_{- \pm \pm}^{m+2} \;_2{\tilde{F}_2}\left(1,\frac{m+2}{2};\frac{3}{2},\frac{m+4}{2};-\left(\frac{r_{- \pm \pm}}{2p}\right)^2 \right) \Bigg] \nonumber \\
    & \qquad = \chi_{+ \pm \pm}^{(m)}(r_1,r_2,r_3;p) - \chi_{- \pm \pm}^{(m)}(r_1,r_2,r_3;p),
\end{align}
where $\Gamma$ is the (complete) Gamma function and $_2{\tilde{F}_2}$ is the regularized hypergeometric function. We have defined
\begin{align}
\label{eqn:chi}
    \chi_{\pm \pm \pm}^{(m)}(r_1,r_2,r_3;p) &\equiv -\frac{i}{4p}\Gamma\left(\frac{m+2}{2}\right)r_{\pm \pm \pm}^{m+2}\;_2{\tilde{F}_2}\left(1,\frac{m+2}{2};\frac{3}{2},\frac{m+4}{2};-\left(\frac{r_{\pm \pm \pm}}{2p}\right)^2 \right).
\end{align}
The first base case can be written in its final form as

{\color{red}
\begin{align}
\label{eqn:gauss_bc1}
    f_{\ell,0,0}(r_1,r_2,r_3;p) &= -\frac{(-i)^{\ell}\sqrt{\pi}}{16pr_2r_3}\frac{1}{2^{\ell}}\sum\limits_{n=0}^{\lfloor \ell/2 \rfloor}(-1)^n\binom{\ell}{n}\binom{2(\ell-n)}{\ell}r_1^{-1-\ell+2n}\sum\limits_{m=0}^{\ell-2n}\binom{\ell-2n}{m} \nonumber \\
    & \qquad \times \Big[(-r_2-r_3)^{\ell-2n-m}\left(G_{-++}^{(m)}-G_{+++}^{(m)}+\chi_{-++}^{(m)}-\chi_{+++}^{(m)} \right) \Big. \nonumber \\
    & \Big. \qquad - (-r_2+r_3)^{\ell-2n-m}\left(G_{-+-}^{(m)}-G_{++-}^{(m)}+\chi_{-+-}^{(m)}-\chi_{++-}^{(m)} \right) \Big. \nonumber \\
    &\Big. \qquad - (r_2-r_3)^{\ell-2n-m}\left(G_{--+}^{(m)}-G_{+-+}^{(m)}+\chi_{--+}^{(m)}-\chi_{+-+}^{(m)} \right) \Big. \nonumber \\
    &\Big. \qquad +(r_2+r_3)^{\ell-2n-m}\left(G_{---}^{(m)}-G_{+--}^{(m)}+\chi_{---}^{(m)}-\chi_{+--}^{(m)} \right) \Big]\textcolor{black}{,}
\end{align}}

where the arguments of $G_{\pm\pm\pm}^{(m)}$ and $\chi_{\pm\pm\pm}^{(m)}$ have been suppressed for legibility.

The other two base cases are obtained by following the same steps outlined throughout this section, resulting in
{\color{red}
\begin{align}
\label{eqn:gauss_bc2}
    f_{\ell,-1,0}(r_1,r_2,r_3;p) &= \frac{(-i)^{\ell+1}\sqrt{\pi}}{16pr_2r_3}\frac{1}{2^{\ell}}\sum\limits_{n=0}^{\lfloor \ell/2 \rfloor}(-1)^n\binom{\ell}{n}\binom{2(\ell-n)}{\ell}r_1^{-1-\ell+2n}\sum\limits_{m=0}^{\ell-2n}\binom{\ell-2n}{m} \nonumber \\
    & \qquad \times \Big[(-r_2-r_3)^{\ell-2n-m}\left(G_{-++}^{(m)}-G_{+++}^{(m)}+\chi_{-++}^{(m)}-\chi_{+++}^{(m)} \right) \Big. \nonumber \\
    & \Big. \qquad - (-r_2+r_3)^{\ell-2n-m}\left(G_{-+-}^{(m)}-G_{++-}^{(m)}+\chi_{-+-}^{(m)}-\chi_{++-}^{(m)} \right) \Big. \nonumber \\
    &\Big. \qquad + (r_2-r_3)^{\ell-2n-m}\left(G_{--+}^{(m)}-G_{+-+}^{(m)}+\chi_{--+}^{(m)}-\chi_{+-+}^{(m)} \right) \Big. \nonumber \\
    &\Big. \qquad - (r_2+r_3)^{\ell-2n-m}\left(G_{---}^{(m)}-G_{+--}^{(m)}+\chi_{---}^{(m)}-\chi_{+--}^{(m)} \right) \Big]
\end{align}}
and 
{\color{red}
\begin{align}
\label{eqn:gauss_bc3}
    f_{\ell,-1,-1}(r_1,r_2,r_3;p) &= \frac{(-i)^{\ell}\sqrt{\pi}}{16pr_2r_3}\frac{1}{2^{\ell}}\sum\limits_{n=0}^{\lfloor \ell/2 \rfloor}(-1)^n\binom{\ell}{n}\binom{2(\ell-n)}{\ell}r_1^{-1-\ell+2n}\sum\limits_{m=0}^{\ell-2n}\binom{\ell-2n}{m} \nonumber \\
    & \qquad \times \Big[(-r_2-r_3)^{\ell-2n-m}\left(G_{-++}^{(m)}-G_{+++}^{(m)}+\chi_{-++}^{(m)}-\chi_{+++}^{(m)} \right) \Big. \nonumber \\
    & \Big. \qquad + (-r_2+r_3)^{\ell-2n-m}\left(G_{-+-}^{(m)}-G_{++-}^{(m)}+\chi_{-+-}^{(m)}-\chi_{++-}^{(m)} \right) \Big. \nonumber \\
    &\Big. \qquad + (r_2-r_3)^{\ell-2n-m}\left(G_{--+}^{(m)}-G_{+-+}^{(m)}+\chi_{--+}^{(m)}-\chi_{+-+}^{(m)} \right) \Big. \nonumber \\
    &\Big. \qquad +(r_2+r_3)^{\ell-2n-m}\left(G_{---}^{(m)}-G_{+--}^{(m)}+\chi_{---}^{(m)}-\chi_{+--}^{(m)} \right) \Big]\textcolor{black}{.}
\end{align}}
We note that the powers of the pairs of $r_i$ serving as pre-factors in each line above are always greater than or equal to zero; $m \leq \ell - 2n$, and $\ell - 2n \geq 0$ due to the floor function in the outer sum (which ultimately stems from expanding the Legendre polynomial in a series as in equation \ref{eqn:leg_pow}). The sign on each pair of $r_i$ stems from writing the $j_0(kr_i)$ and $j_{-1}(kr_i)$ in terms of sine or cosine, then rewriting as complex exponentials.

Given these three base cases, the recursion relation (\ref{eqn:int_recursion}) may now be used to construct the integral (\ref{eqn:gauss_rec}) for any non-negative integer values of the $\ell_i$. As in \S \ref{sssec:exp_rec}, the pre-factors of equations (\ref{eqn:gauss_bc1}), (\ref{eqn:gauss_bc2}), and (\ref{eqn:gauss_bc3}) as well as the signs on each line differ because of the multiplication of sines and cosines as complex exponentials.

\section{Stability of the Recursion Relations}
\label{sec:stability}
For many applications within cosmology, such as covariance matrix calculations, the orders of two of the sBFs within a triple-sBF integral each have a maximum value $\leq 10$ (\textit{e.g.} \cite{sarabande, hou_parity, 3pt_alg, jiamin_covar, encore}). The maximum value of the order of the third sBF is set by the sum of the orders of the other two sBFs; this is because of the triangle inequalities that stem from Wigner $3$-j symbols coupling the sBF orders. Thus, it is desirable for the recursion relations of \S \ref{sssec:exp_rec} and \S \ref{sec:gauss_rec} to be stable for $\ell \leq 20$.

We duplicate the recursion for triple-sBF integrals (equation \ref{eqn:int_recursion}) below:
\begin{align}
\label{eqn:recursion_dup}
    f_{\ell_1,\ell_2+1,\ell_3} &= \left(\frac{r_1}{r_2}
\right)\left(\frac{2\ell_2+1}{2\ell_1+1}\right)\left(f_{\ell_1-1,\ell_2,\ell_3} + f_{\ell_1+1,\ell_2,\ell_3}\right) - f_{\ell_1,\ell_2-1,\ell_3}.
\end{align}
$f_{\ell,0,0}$, $f_{\ell,-1,0}$, and $f_{\ell,-1,-1}$ are the three base cases needed to anchor this recursion.

We will use an example to show how the integrals with higher-order sBFs are sums of integrals with lower-order sBFs. We begin by evaluating one of the lowest-order $f$-integrals that is not one of the base cases: $f_{1,1,0}$. We use the recursion (equation \ref{eqn:recursion_dup}) to write $f_{1,1,0}$ as
\begin{align}
\label{eqn:f110_rec}
    f_{1,1,0} &= \left(\frac{r_1}{r_2}\right)\left(\frac{1}{3}\right)\left(f_{0,0,0}+f_{2,0,0} \right) - f_{1,-1,0}.
\end{align}
The $f_{0,0,0}$, $f_{2,0,0}$, and $f_{1,-1,0}$ in equation (\ref{eqn:f110_rec}) are all base cases. Thus, we see that $f_{1,1,0}$ can be evaluated by first determining only three base cases.

We now consider $f_{1,1,1}$, an $f$-integral with higher sBF orders than $f_{1,1,0}$. Using the recursion (equation \ref{eqn:recursion_dup}), $f_{1,1,1}$ is 
\begin{align}
\label{eqn:f111_rec}
    f_{1,1,1} &= \left(\frac{r_1}{r_2}\right)\left(\frac{1}{3}\right)\left(f_{0,0,1}+\textcolor{red}{f_{2,0,1}} \right) - \textcolor{blue}{f_{1,-1,1}}.
\end{align}
In equation (\ref{eqn:f111_rec}), $f_{0,0,1}$ is a base case; $f_{2,0,1}$ and $f_{1,-1,1}$ are not base cases and will each need to be evaluated in terms of base cases by using the recursion. 

$f_{2,0,1}$ is the same as $f_{2,1,0}$; switching the ordering of the sBFs within a triple-sBF integral does not affect the integral. We thus have
\begin{align}
\label{eqn:f210_rec}
    \textcolor{red}{f_{2,0,1}} = f_{2,1,0} =  \left(\frac{r_1}{r_3}\right)\left(\frac{1}{5}\right)\left(f_{1,0,0}+f_{3,0,0} \right) - f_{2,-1,0}.
\end{align}
The $f_{1,0,0}$, $f_{3,0,0}$, and $f_{2,-1,0}$ in equation (\ref{eqn:f210_rec}) are all base cases. The factor $(r_1/r_2)$ in equation (\ref{eqn:recursion_dup}) has been replaced by $(r_1/r_3)$ in equation (\ref{eqn:f210_rec}) since we have switched the ordering of the sBFs with arguments $(kr_2)$ and $(kr_3)$ when evaluating $f_{2,1,0}$ instead of $f_{2,0,1}$.

Finally, we must use the recursion to evaluate the $f_{1,-1,1}$ (equal to $f_{1,1,-1}$) in equation (\ref{eqn:f111_rec}):
\begin{align}
\label{eqn:f1_minus1_1_rec}
    \textcolor{blue}{f_{1,-1,1}} = f_{1,1,-1} = \left(\frac{r_1}{r_3}\right)\left(\frac{1}{3}\right)\left(f_{0,0,-1}+f_{2,0,-1} \right) - f_{1,-1,-1}.
\end{align}
In equation (\ref{eqn:f1_minus1_1_rec}), the $f_{0,0,-1}$, $f_{2,0,-1}$, and $f_{1,-1,-1}$ are all base cases. Again, the factor $(r_1/r_3)$ in equation (\ref{eqn:f1_minus1_1_rec}) arises from swapping the ordering of two sBFs within the desired integral.

We now see that $f_{1,1,1}$ (equation \ref{eqn:f111_rec}) depends on seven base cases: one is $f_{0,0,1}$ (equation \ref{eqn:f111_rec}), three are from $f_{2,0,1}$ (equation \ref{eqn:f210_rec}), and three are from $f_{1,-1,1}$ (equation \ref{eqn:f1_minus1_1_rec}). This example shows that non-base case integrals are simply sums over some number of base case integrals. For example, the lowest order sBF to be used in the recursion is $\ell_1 = -1$. If $\ell_2$ and $\ell_3$ are 0 or -1, the recursion is simply composed of base cases. By using the recursion along with these base cases, any triple sBF integral in the plane $\ell_1 = -1$ with arbitrary $\{\ell_2,\ell_3\}$ can be evaluated. This same strategy can be used for any other $\ell_1 > -1$ plane: first evaluate the base cases when $\ell_2$ and $\ell_3$ are 0 or 1, then use the recursion to build up to integrals with higher-order $\ell_2$ and $\ell_3$. Since swapping the order of any of the sBFs within a triple-sBF integral does not alter the integral, the same recursive technique can be used to evaluate integrals in the planes $\ell_2 \geq -1$ and $\ell_3 \geq -1$.

The stability of the recursion depends on whether a small error present in the base cases will propagate with each iteration of the recursion. As we use the recursion to evaluate higher-order triple-sBF integrals, we are simply adding more terms. Each new term may have a random error $\epsilon$, which is independent of previous terms as it represents numerical error from evaluating the special functions that are required for the base cases. When adding terms in the recursion, the error from each will add up in quadrature, as is typical for a random walk. Thus, the stability of the recursion can be viewed as the expected error after $N$ iterations of the recursion, which is $\sqrt{N}\epsilon$. 

To practically use the recursive work of this paper, we must address two issues: accurately evaluating each base case and co-adding as many base cases as needed without loss of numerical precision. This second issue is widespread in numerical work with many remedies for it. For instance, the \textsc{mp.math} library in \textsc{Python} offers arbitrary precision as a custom type. Balanced sum algorithms also exist to handle addition of values that significantly differ in order of magnitude, as could occur when co-adding a number of the base cases. Although it is possible to accurately evalulate the base cases, underflow may be present in the computed values. This underflow will build up throughout the recursion and can make the recursion unstable if high-precision results are needed. Additionally, sBF integrals solved using the spherical harmonic shift theorem (as in \cite{mehrem_londergan}) have a known numerical issue when one of the free arguments of the sBFs is much less than the other two free arguments. However, our approach does not require this shift theorem, so we do not expect to see any numerical issues with our results, similar to the explanation in \cite{mehrem_londergan} that the integrals within \cite{jackson1972integrals} do not have any numerical issues. We now turn to the first issue of accurately evaluating the base cases.

\subsection{Stability of Recursion with Exponential Damping}
\label{ssec:exp_stability}
The base cases (equations \ref{eqn:exp_bc1}, \ref{eqn:exp_bc2}, and \ref{eqn:exp_bc3}) for the exponentially-damped triple-sBF integral (\S \ref{sssec:exp_rec}) depend on Legendre functions of the second kind $Q_{\ell}(R_{\pm\pm})$, with $R_{\pm\pm} \equiv (-ip^2 \pm r_2 \pm r_3)/r_1$ previously defined in equation (\ref{eqn:R_def}). These Legendre functions can be written in terms of the hyperbolic arctangent and powers of the argument $R_{\pm\pm}$ (\cite{GR} equation 8.827), which will diverge when $R_{\pm\pm}$ is along the real line from -1 to +1. Since $R_{\pm\pm}$ is manifestly complex, $Q_{\ell}(R_{\pm\pm})$ has no divergence. 

We display the five lowest-order Legendre functions below:
\begin{align}
    Q_0(R_{\pm\pm}) &= \text{tanh}^{-1}(R_{\pm\pm}), \nonumber \\
    Q_1(R_{\pm\pm}) &= R_{\pm\pm}\text{tanh}^{-1}(R_{\pm\pm})-1, \nonumber \\
    Q_2(R_{\pm\pm}) &= \frac{1}{2}\left(3R_{\pm\pm}^2-1 \right)\text{tanh}^{-1}(R_{\pm\pm})-\frac{3}{2}R_{\pm\pm}, \nonumber \\
    Q_3(R_{\pm\pm}) &= \frac{1}{2}\left(5R_{\pm\pm}^3-3R_{\pm\pm} \right)\text{tanh}^{-1}(R_{\pm\pm})-\frac{5}{2}R_{\pm\pm}^2+\frac{2}{3}, \nonumber \\
    Q_4(R_{\pm\pm}) &= \frac{1}{8}\left(35R_{\pm\pm}^4-30R_{\pm\pm}^2+3\right)\text{tanh}^{-1}(R_{\pm\pm})-\frac{35}{8}R_{\pm\pm}^3+\frac{55}{24}R_{\pm\pm}. 
\end{align}
The hyperbolic arctangent is standardly implemented in numerical libraries such as \textsc{scipy}. Thus, the base cases for exponentially-damped triple-sBF integrals are stable.

\subsection{Stability of Recursion with Gaussian Damping}
\label{ssec:gauss_stability}
The Gaussian-damped triple-sBF integral (\S \ref{sec:gauss_rec}) has base cases (equations \ref{eqn:gauss_bc1}, \ref{eqn:gauss_bc2}, and \ref{eqn:gauss_bc3}) that depend on $G_{\pm\pm\pm}^{(m)}$ (equation \ref{eqn:G}) and $\chi_{\pm\pm\pm}^{(m)}$ (equation \ref{eqn:chi}), duplicated below:
\begin{align}
\label{eqn:G_duplicate}
    G_{\pm \pm \pm}^{(m)}(r_1,r_2,r_3;p) &\equiv 2^mp^{m+1}\Gamma\left(\frac{m+1}{2},\left(\frac{r_{\pm \pm \pm}}{2p}\right)^2 \right) \\
\end{align}
and
\begin{align}
\label{eqn:chi_duplicate}
    \chi_{\pm \pm \pm}^{(m)}(r_1,r_2,r_3;p) &\equiv -\frac{i}{4p}\Gamma\left(\frac{m+2}{2}\right)r_{\pm \pm \pm}^{m+2}\;_2{\tilde{F}_2}\left(1,\frac{m+2}{2};\frac{3}{2},\frac{m+4}{2};-\left(\frac{r_{\pm \pm \pm}}{2p}\right)^2 \right).
\end{align}
We have previously defined $r_{\pm\pm\pm} \equiv \pm r_1 \pm r_2 \pm r_3$ in equation (\ref{eqn:r_def}). Accurate numerical algorithms exist for the incomplete gamma function in $G_{\pm \pm \pm}^{(m)}$ (equation \ref{eqn:G_duplicate}) \cite{eval_inc_gamma, comp_inv_inc_gamma, comp_inc_gamma} and the hypergeometric function in $\chi_{\pm \pm \pm}^{(m)}$ (equation \ref{eqn:chi_duplicate}) \cite{uni_comp_hyp, hyp_geometric}. The base cases for the Gaussian-damped recursion are therefore stable.

\section{Damped Triple-sBF Integrals Weighted by Higher Powers of k}
\label{sec:para_diff}
In the preceding sections, we evaluated recursion relations for damped triple-sBF integrals involving the Jacobian $k^2dk$. It was necessary to begin with $k^2$ so that the sBFs with orders 0 and -1 in the base cases, when rewritten as sines and cosines divided by their arguments $kr_i$, would cancel the Jacobian. Recursions for triple-sBF integrals involving higher powers of $k$ cannot be evaluated using the methods described in \S \ref{sssec:exp_rec} and \S \ref{sec:gauss_rec}; however, they can be obtained by parametric differentiation.

We begin with the triple-sBF integral with exponential damping given by equation (\ref{eqn:exp_general}):
\begin{align}
    f_{\ell_1,\ell_2, \ell_3}(r_1,r_2,r_3;p) &= \int_{0}^{\infty}k^2dk\;e^{-p^2k}j_{\ell_1}(kr_1)j_{\ell_2}(kr_2)j_{\ell_3}(kr_3),
\end{align}
for $\{p, \ell_i, r_i\} \in \mathbb{R}$, $\ell_i \in \mathbb{Z}$, $\ell_i \geq 0$, $r_1 > 0$, $r_i \neq 0$ and $p \neq 0$. The partial derivative of $f_{\ell_1,\ell_2, \ell_3}(r_1,r_2,r_3;p)$ with respect to $-p^2$ can be moved inside the integral since the derivative is with respect to $-p^2$ and the integral is with respect to $k$:
\begin{align}
    \frac{\partial}{\partial (-p^2)}\left(f_{\ell_1,\ell_2, \ell_3}(r_1,r_2,r_3;p)\right) &= \int_{0}^{\infty} \frac{\partial}{\partial (-p^2)} k^2dk\;e^{-p^2k}j_{\ell_1}(kr_1)j_{\ell_2}(kr_2)j_{\ell_3}(kr_3) \nonumber \\
    &= \int_{0}^{\infty}k^3dk\;e^{-p^2k}j_{\ell_1}(kr_1)j_{\ell_2}(kr_2)j_{\ell_3}(kr_3),
\end{align}
where the partial derivative has resulted in one extra power of $k$ in the integral since the only $-p^2$ dependence is in the exponential. Each successive derivative will add one to the power of $k$; therefore, this can be generalized to 
\begin{align}
    \frac{\partial^n}{\partial (-p^2)^n}\left(f_{\ell_1,\ell_2, \ell_3}(r_1,r_2,r_3;p)\right) &= \int_{0}^{\infty}k^{2+n}dk\;e^{-p^2k}j_{\ell_1}(kr_1)j_{\ell_2}(kr_2)j_{\ell_3}(kr_3).
\end{align}
We can therefore evaluate the recursion for triple-sBF integrals with exponential damping for $k$ to any power $\geq 2$ by first evaluating the recursion for $k^2$ as done in \S \ref{sssec:exp_rec} and then taking a partial derivative of the result with respect to $-p^2$ for each additional power of $k$ desired.

This parametric differentiation method can also be used to evaluate triple-sBF integrals with Gaussian damping involving $k$ to any even power $\geq 2$. We begin with equation (\ref{eqn:gauss_rec}):
\begin{align}
    f_{\ell_1,\ell_2, \ell_3}(r_1,r_2,r_3;p) &= \int_{0}^{\infty}k^2dk\;e^{-(pk)^2}j_{\ell_1}(kr_1)j_{\ell_2}(kr_2)j_{\ell_3}(kr_3),
\end{align}
for $\{p, \ell_i, r_i\} \in \mathbb{R}$, $\ell_i \in \mathbb{Z}$, $\ell_i \geq 0$, $r_1 > 0$, $r_i \neq 0$, and $p \neq 0$. The partial derivative of $f_{\ell_1,\ell_2, \ell_3}(r_1,r_2,r_3;p)$ with respect to $-p^2$ can be moved inside the integral since the derivative is with respect to $-p^2$ while the integral is with respect to $k$. Since the only $-p^2$ dependence is in the Gaussian, taking this derivative will increase the power of $k$ by two:
\begin{align}
    \frac{\partial}{\partial (-p^2)}\left(f_{\ell_1,\ell_2, \ell_3}(r_1,r_2,r_3;p)\right) &= \int_{0}^{\infty} \frac{\partial}{\partial (-p^2)} k^2dk\;e^{-(pk)^2}j_{\ell_1}(kr_1)j_{\ell_2}(kr_2)j_{\ell_3}(kr_3) \nonumber \\
    &= \int_{0}^{\infty}k^{4}dk\;e^{-(pk)^2}j_{\ell_1}(kr_1)j_{\ell_2}(kr_2)j_{\ell_3}(kr_3).
\end{align}
This can be generalized to
\begin{align}
    \frac{\partial^n}{\partial (-p^2)^n}\left(f_{\ell_1,\ell_2, \ell_3}(r_1,r_2,r_3;p)\right) &= \int_{0}^{\infty}k^{2+2n}dk\;e^{-(pk)^2}j_{\ell_1}(kr_1)j_{\ell_2}(kr_2)j_{\ell_3}(kr_3),
\end{align}
since each successive derivative with respect to $-p^2$ will increase the power of $k$ by two. By evaluating the recursion for triple-sBF integrals with Gaussian damping involving the Jacobian $k^2dk$ (\S \ref{sec:gauss_rec}) and then taking the necessary amount of partial derivatives with respect to $-p^2$ of the result, we can evaluate triple-sBF integrals with Gaussian damping weighted by $k$ to any even power $\geq 2$.

\section{A Second Method to Evaluate Triple-sBF Integrals with Gaussian Damping}
\label{sec:bowman}
We now show an alternate, non-recursive method to evaluate triple-sBF integrals with Gaussian damping weighted by any non-negative integer power of $k$. The result will depend upon incomplete Gamma functions and hypergeometric functions. We term this the ``Hankel-Bowman'' method since it uses a result from Bowman \cite{bowman} derived from Hankel's contour integral. 

We begin with
\begin{align}
\label{eqn:bowman_orig}
   f_{\ell_1,\ell_2,\ell_3}(r_1,r_2,r_3;p) = \int_{0}^{\infty}k^ndk\;e^{-(pk)^2}j_{\ell_1}(kr_1)j_{\ell_2}(kr_2)j_{\ell_3}(kr_3)
\end{align} for $\{p, \ell_i, r_i\} \in \mathbb{R}$, $\{\ell_i,n\} \in \mathbb{Z}$, $\ell_i \geq 0$, $r_i \neq 0$, $p \neq 0$ and $n \geq 0$. The spherical Bessel functions are related to cylindrical Bessel functions $J_{\ell}(x)$ by 
\begin{align}
\label{eqn:sph_to_cyl}
    j_{\ell}(x) = \sqrt{\frac{\pi}{2x}}J_{\ell+1/2}(x).
\end{align}
Hankel's contour integral \cite{bowman} can be used to rewrite the cylindrical Bessel functions as
\begin{align}
\label{eqn:hankel_contour}
    J_{\ell}(x) = \frac{x^{\ell}}{2^{\ell}\sqrt{\pi}\;\Gamma(\ell+\frac{1}{2})}\int_{-1}^{1}dq\;e^{ixq}(1-q^2)^{\ell-\frac{1}{2}}
\end{align}
for $\ell > -(1/2).$
We first use equation (\ref{eqn:sph_to_cyl}) to write each sBF in equation (\ref{eqn:bowman_orig}) as a cylindrical Bessel function, then use equation (\ref{eqn:hankel_contour}) to rewrite each cylindrical Bessel function as an integral over $q_i$, and change the order of integration so that the $k$ integral will be performed first:
\begin{align}
\label{eqn:f_4ints}
    f_{\ell_1,\ell_2, \ell_3}(r_1,r_2,r_3;p) &= \frac{2^{-L-3}}{\Gamma(\ell_1+1)\Gamma(\ell_2+1)\Gamma(\ell_3+1)}\;r_1^{\ell_1}r_2^{\ell_2}r_3^{\ell_3}\int_{-1}^1dq_1\;(1-q_1^2)^{\ell_1} \nonumber \\
    & \qquad \times\int_{-1}^1dq_2\;(1-q_2^2)^{\ell_2}\int_{-1}^1dq_3\;(1-q_3^2)^{\ell_3}\int_{0}^{\infty}dk\;k^{n+L}e^{-(pk)^2+iks_{123}},
\end{align}
where we have defined $L \equiv \ell_1+\ell_2+\ell_3$ and 
\begin{align}
\label{eqn:s123}
    s_{123} \equiv r_1q_1+r_2q_2+r_3q_3.
\end{align}

The inner $k$ integral, which will be defined as $I_k$ below, will be performed first:
\begin{align}
\label{eqn:bowman_Ik}
    I_k(r_1,r_2,r_3;p) &\equiv \int_{0}^{\infty}dk\;k^{n+L}e^{-(pk)^2+iks_{123}} \nonumber \\
    &= e^{-[s_{123}/(2p)]^2}\int_{0}^{\infty}dk\;k^{n+L}\exp\left[-p^2\left(k-\frac{is_{123}}{2p^2}\right)^2\right] \nonumber\\
    &= e^{-[s_{123}/(2p)]^2}\int_{(-is_{123})/(2p^2)}^{\infty}dx\;\left(x+\frac{is_{123}}{2p^2}\right)^{n+L}e^{-(px)^2} \nonumber\\
    &= e^{-[s_{123}/(2p)]^2}\sum\limits_{\zeta=0}^{n+L}\binom{n+L}{\zeta}\left(\frac{is_{123}}{2p^2}\right)^\zeta\int_{(-is_{123})/(2p^2)}^{\infty}dx\;x^{n+L-\zeta}e^{-(px)^2}.
\end{align}
We first complete the square in the Gaussian (to obtain the second line of equation \ref{eqn:bowman_Ik}), then use a change of variable $x = k-[(is_{123})/(2p^2)]$ (to obtain the third line), and finally use a binomial expansion on the polynomial within the integrand (to obtain the fourth line). The integral in the last line of equation (\ref{eqn:bowman_Ik}) is known in closed form in terms of an incomplete gamma function; we thus obtain\footnote{https://functions.wolfram.com/ElementaryFunctions/Exp/21/01/02/01/01/08/0001/}
\begin{align}
\label{eqn:bowman_Ik_final}
    I_{k}(r_1,r_2,r_3;p) &= \frac{1}{2}e^{-[s_{123}/(2p)]^2}\sum\limits_{\zeta=0}^{n+L}\binom{n+L}{\zeta}\left(\frac{is_{123}}{2p^2}\right)^\zeta p^{-(\alpha+1)}\Gamma\left(\frac{\alpha+1}{2},-\left(\frac{s_{123}}{2p} \right)^2 \right).
\end{align}
We have defined $\alpha \equiv n+L-\zeta$.
Replacing the innermost, $k$ integral in equation (\ref{eqn:f_4ints}) using equation (\ref{eqn:bowman_Ik_final}), equation (\ref{eqn:f_4ints}) becomes
\begin{align}
\label{eqn:f_inc_gamma}
    f_{\ell_1,\ell_2, \ell_3}(r_1,r_2,r_3;p) &= \frac{2^{-L-4}}{\Gamma(\ell_1+1)\Gamma(\ell_2+1)\Gamma(\ell_3+1)}\;r_1^{\ell_1}r_2^{\ell_2}r_3^{\ell_3}\sum\limits_{\zeta=0}^{n+L}\binom{n+L}{\zeta}\left(\frac{i}{2p^2}\right)^\zeta p^{-(\alpha+1)}
    \nonumber\\
    & \qquad \times \int_{-1}^1dq_1\;(1-q_1^2)^{\ell_1}\int_{-1}^1dq_2\;(1-q_2^2)^{\ell_2}\int_{-1}^1dq_3\;(1-q_3^2)^{\ell_3}e^{-[s_{123}/(2p)]^2}\nonumber\\
    &\qquad\qquad\qquad\qquad\qquad\qquad\qquad\qquad\qquad \times \Gamma\left(\frac{\alpha+1}{2},-\left(\frac{s_{123}}{2p} \right)^2 \right)s_{123}^{\zeta}.
\end{align}

The upper incomplete gamma function is the difference between the complete gamma function and the lower incomplete gamma function, $\gamma$:\footnote{https://mathworld.wolfram.com/IncompleteGammaFunction.html}
\begin{align}
\label{eqn:upper_gamma}
    \Gamma\left(\frac{\alpha+1}{2},-\left(\frac{s_{123}}{2p} \right)^2 \right) &= \Gamma\left(\frac{\alpha+1}{2}\right) - \gamma\left(\frac{\alpha+1}{2},-\left(\frac{s_{123}}{2p} \right)^2\right).
\end{align}
The lower incomplete gamma function can be rewritten in terms of the confluent hypergeometric function of the first kind so that equation (\ref{eqn:upper_gamma}) becomes\footnote{https://mathworld.wolfram.com/IncompleteGammaFunction.html}
\begin{align}
\label{eqn:lower_gamma}
    \Gamma\left(\frac{\alpha+1}{2},-\left(\frac{s_{123}}{2p} \right)^2 \right) &= \Gamma\left(\frac{\alpha+1}{2}\right) - \frac{2}{\alpha+1}(-1)^{(\alpha+1)/2}\left(\frac{s_{123}}{2p}\right)^{\alpha+1}e^{[s_{123}/(2p)]^2}\nonumber \\
    & \qquad \times \;_1{F_1}\left(1;\frac{\alpha+3}{2}; -\left(\frac{s_{123}}{2p} \right)^2 \right).
\end{align}

After replacing the upper incomplete gamma function in equation (\ref{eqn:f_inc_gamma}) using equation (\ref{eqn:lower_gamma}), we obtain
\begin{align}
\label{eqn:f_in_terms_of_AB}
    f_{\ell_1,\ell_2, \ell_3}(r_1,r_2,r_3;p) &= \frac{2^{-L-4}}{\Gamma(\ell_1+1)\Gamma(\ell_2+1)\Gamma(\ell_3+1)}\;r_1^{\ell_1}r_2^{\ell_2}r_3^{\ell_3}\sum\limits_{\zeta=0}^{n+L}\binom{n+L}{\zeta}\left(\frac{i}{2p^2}\right)^\zeta p^{-(\alpha+1)}
    \nonumber\\
    & \qquad \times \int_{-1}^1dq_1\;(1-q_1^2)^{\ell_1}\int_{-1}^1dq_2\;(1-q_2^2)^{\ell_2}\int_{-1}^1dq_3\;(1-q_3^2)^{\ell_3}s_{123}^\zeta \nonumber \\ 
    & \qquad \times\left[\Gamma\left(\frac{\alpha+1}{2}\right)e^{-[s_{123}/(2p)]^2} \right. \nonumber \\
    & \left. \qquad\qquad -\frac{2}{\alpha+1}(-1)^{(\alpha+1)/2}(2p)^{-(\alpha+1)}s_{123}^{\alpha+1} \;_1{F_1}\left(1;\frac{\alpha+3}{2}; -\left(\frac{s_{123}}{2p} \right)^2 \right) \right] \nonumber \\
    &= \frac{2^{-L-4}}{\Gamma(\ell_1+1)\Gamma(\ell_2+1)\Gamma(\ell_3+1)}\;r_1^{\ell_1}r_2^{\ell_2}r_3^{\ell_3}\sum\limits_{\zeta=0}^{n+L}\binom{n+L}{\zeta}\left(\frac{i}{2p^2}\right)^\zeta p^{-(\alpha+1)} \nonumber \\
    & \qquad \times (I_{\text{pl}}-I_{\text{hg}}),
\end{align}
where we have defined two new integrals
\begin{align}
\label{eqn:IA_orig}
    I_{\text{pl}}(r_1,r_2,r_3;p) &= \Gamma\left(\frac{\alpha+1}{2}\right)\int_{-1}^1dq_1\;(1-q_1^2)^{\ell_1}\int_{-1}^1dq_2\;(1-q_2^2)^{\ell_2}\int_{-1}^1dq_3\;(1-q_3^2)^{\ell_3}e^{-[s_{123}/(2p)]^2} \nonumber \\
    &\qquad\qquad\qquad\qquad\qquad\qquad\qquad\qquad\qquad\qquad\qquad\qquad\qquad\times s_{123}^\zeta
\end{align}
and
\begin{align}
\label{eqn:IB_orig}
    I_{\text{hg}}(r_1,r_2,r_3;p) &= \frac{2}{\alpha+1}(-1)^{(\alpha+1)/2}(2p)^{-(\alpha+1)}\int_{-1}^1dq_1\;(1-q_1^2)^{\ell_1}\int_{-1}^1dq_2\;(1-q_2^2)^{\ell_2}\nonumber \\
    &\qquad\qquad\qquad\qquad\qquad\times \int_{-1}^1dq_3\;(1-q_3^2)^{\ell_3}s_{123}^{n+L+1}\;_1{F_1}\left(1;\frac{\alpha+3}{2}; -\left(\frac{s_{123}}{2p} \right)^2 \right).
\end{align}
$I_{\text{pl}}$ depends on a power law and $I_{\text{hg}}$ depends on a hypergeometric function.
To obtain equation (\ref{eqn:IB_orig}), we used $\zeta+\alpha+1 = n+L+1$.
\\
\indent We begin with the integral $I_{\text{pl}}$. We need to recall that $s_{123}$ depends on $q_3$ (equation \ref{eqn:s123}). To evaluate the $q_3$ integral in equation (\ref{eqn:IA_orig}), we make the change of variable $x = -[s_{123}/(2p)]^2$, then use $q_3 = [2ip\sqrt{x}-r_1q_1-r_2q_2]/r_3$ to rewrite $(1-q_3^2)^{\ell_3}$ in terms of $x$  and use the binomial expansion on it:
\begin{align}
\label{eqn:IA_3sub}
    I_{\text{pl}}(r_1,r_2,r_3;p) &= \Gamma\left(\frac{\alpha+1}{2}\right)\frac{ip}{r_3}(2ip)^\zeta \sum\limits_{b=0}^{\ell_3}\binom{\ell_3}{b}(-1)^br_3^{-2b}\sum\limits_{c=0}^{2b}\binom{2b}{c}(2ip)^{2b-c}(-1)^c \nonumber \\
    & \qquad \times \int_{-1}^1dq_1\;(1-q_1^2)^{\ell_1}\int_{-1}^1dq_2\;(1-q_2^2)^{\ell_2}s_{12}^c\int_{-[(s_{12}-r_3)/(2p)]^2}^{-[(s_{12}+r_3)/(2p)]^2}dx\;x^{\beta}e^x,
\end{align}
where we have defined $\beta \equiv [\zeta+2b-c-1]/2$ and
\begin{align}
\label{eqn:s12}
    s_{12} \equiv r_1q_1+r_2q_2.
\end{align}
We evaluate the innermost integral with respect to $x$ using \cite{GR} equation 2.33.10. Equation (\ref{eqn:IA_3sub}) then becomes
\begin{align}
\label{eqn:IA_3}
    I_{\text{pl}}(r_1,r_2,r_3;p) &= \Gamma\left(\frac{\alpha+1}{2}\right)\frac{ip}{r_3}(2ip)^\zeta \sum\limits_{b=0}^{\ell_3}\binom{\ell_3}{b}(-1)^br_3^{-2b}\sum\limits_{c=0}^{2b}\binom{2b}{c}(2ip)^{2b-c}(-1)^{c-\beta} \nonumber \\
    & \qquad \times \int_{-1}^1dq_1\;(1-q_1^2)^{\ell_1}\int_{-1}^1dq_2\;(1-q_2^2)^{\ell_2}s_{12}^c\left[\Gamma\left(\beta+1,\left(\frac{s_{12}+r_3}{2p} \right)^2 \right) \right. \nonumber \\
    & \qquad\qquad\qquad\qquad\qquad\qquad\qquad\qquad\qquad\;\;\;\;\; \left. - \Gamma\left(\beta+1,\left(\frac{s_{12}-r_3}{2p} \right)^2 \right) \right].
\end{align}
\\
\indent To evaluate the innermost, $q_2$ integral, we make the change of variable $x = [(s_{12}\pm r_3)/(2p)]^2$, where the $\pm$ sign represents the sign of $r_3$ in the arguments of the incomplete gamma functions in equation (\ref{eqn:IA_3}). After making this change of variable, we use a binomial expansion on the resulting factor of $s_{12}^c = (2p\sqrt{x}\mp r_3)^c$. We use $q_2 = [2p\sqrt{x}-r_1q_1 \mp r_3]/r_2$ to rewrite $(1-q_2^2)^{\ell_2}$, then use another binomial expansion on this factor to obtain
\begin{align}
\label{eqn:IA_2sub}
    I_{\text{pl}}(r_1,r_2,r_3;p) &= \Gamma\left(\frac{\alpha+1}{2}\right)\frac{ip^2}{r_2r_3}(2ip)^\zeta \sum\limits_{b=0}^{\ell_3}\binom{\ell_3}{b}(-1)^br_3^{-2b}\sum\limits_{c=0}^{2b}\binom{2b}{c}(2ip)^{2b-c}(-1)^{c-\beta}\sum\limits_{d=0}^{\ell_2}\binom{\ell_2}{d} \nonumber \\
    & \qquad \times (-1)^dr_2^{-2d}\sum\limits_{g=0}^{2d}\binom{2d}{g}(2p)^{2d-g}\sum\limits_{h=0}^c\binom{c}{h}(2p)^{c-h}r_3^h\int_{-1}^1dq_1\;(1-q_1^2)^{\ell_1} \nonumber \\
    & \qquad \times \left[(-1)^h(-r_1q_1-r_3)^g\int_{[r_{q-+}/(2p)]^2}^{[r_{q++}/(2p)]^2}dx\;x^{\epsilon}\Gamma\left(\beta+1, x\right) \right. \nonumber \\ 
    &\left. \qquad -(-r_1q_1+r_3)^g\int_{[r_{q--}/(2p)]^2}^{[r_{q+-}/(2p)]^2}dx\;x^{\epsilon}\Gamma\left(\beta+1, x\right) \right],
\end{align}
where we have defined $\epsilon \equiv [c+2d-g-h-1]/2$ and
\begin{align}
    r_{q \pm \pm} \equiv r_1q_1 \pm r_2 \pm r_3.
\end{align}
We now integrate over $x$, resulting in\footnote{https://functions.wolfram.com/GammaBetaErf/Gamma2/21/01/02/01/0001/}
\begin{align}
\label{eqn:IA_2}
    I_{\text{pl}}(r_1,r_2,r_3;p) &= \Gamma\left(\frac{\alpha+1}{2}\right)\frac{ip^2}{r_2r_3}(2ip)^\zeta \sum\limits_{b=0}^{\ell_3}\binom{\ell_3}{b}(-1)^br_3^{-2b}\sum\limits_{c=0}^{2b}\binom{2b}{c}(2ip)^{2b-c}(-1)^{c-\beta}\sum\limits_{d=0}^{\ell_2}\binom{\ell_2}{d} \nonumber \\
    & \qquad \times (-1)^dr_2^{-2d}\sum\limits_{g=0}^{2d}\binom{2d}{g}(2p)^{2d-g}\sum\limits_{h=0}^c\binom{c}{h}(2p)^{c-h}r_3^h\frac{1}{\epsilon+1}\int_{-1}^1dq_1\;(1-q_1^2)^{\ell_1} \nonumber \\
    & \qquad \times \Big\{(-1)^h(-r_1q_1-r_3)^g\left(A_{++}-B_{++}-A_{-+}+B_{-+} \right) \Big. \nonumber\\
    &\qquad\qquad\qquad\qquad\qquad\Big. -(-r_1q_1+r_3)^g \left(A_{+-}-B_{+-}-A_{--}+B_{--} \right) \Big\},
\end{align}
where we have defined (with arguments suppressed)
\begin{align}
    A_{\pm\pm} &\equiv \left(\frac{r_{q \pm \pm}}{2p}\right)^{2(\epsilon+1)}\Gamma\left(\beta+1,\left(\frac{r_{q \pm \pm}}{2p} \right)^2 \right)
\end{align}
and
\begin{align}
    B_{\pm\pm} &\equiv  \Gamma\left(\beta+\epsilon+2,\left(\frac{r_{q \pm \pm}}{2p} \right)^2 \right).
\end{align}
\indent To evaluate the $q_1$ integral of equation (\ref{eqn:IA_2}), we make the change of variable $x = [r_{q \pm \pm}/(2p)]^2$, which corresponds to the arguments of the incomplete gamma functions in equation (\ref{eqn:IA_2}). We use $q_1 = [2p\sqrt{x} \mp r_2 \mp r_3]/r_1$ to rewrite $(1-q_1^2)^{\ell_1}$, then use a binomial expansion on it. We use another binomial expansion on the factor of $(-r_1q_1 \mp r_3)^g$ in the third and fourth lines of equation (\ref{eqn:IA_2}):
\begin{align}
\label{eqn: IA_1sub}
    I_{\text{pl}}(r_1,r_2,r_3;p) &= \Gamma\left(\frac{\alpha+1}{2}\right)\frac{ip^3}{r_1r_2r_3}(2ip)^\zeta \sum\limits_{b=0}^{\ell_3}\binom{\ell_3}{b}(-1)^br_3^{-2b}\sum\limits_{c=0}^{2b}\binom{2b}{c}(2ip)^{2b-c}(-1)^{c-\beta}\sum\limits_{d=0}^{\ell_2}\binom{\ell_2}{d} \nonumber \\
    & \qquad \times (-1)^dr_2^{-2d}\sum\limits_{g=0}^{2d}\binom{2d}{g}(2p)^{2d-g}\sum\limits_{h=0}^c\binom{c}{h}(2p)^{c-h}r_3^h\frac{1}{\epsilon+1}\sum\limits_{m=0}^{\ell_1}\binom{\ell_1}{m}(-1)^mr_1^{-2m} \nonumber \\
    & \qquad \times \sum\limits_{t=0}^{2m}\binom{2m}{t}(2p)^{2m-t}\sum\limits_{u=0}^{g}\binom{g}{u}(-1)^{g-u}r_3^u \sum\limits_{v=0}^{g-u}\binom{g-u}{v}(2p)^{g-u-v} \nonumber \\
    & \qquad \times \left[(-1)^{h+u}(-r_2-r_3)^{t+v}\int_{[r_{-++}/(2p)]^2}^{[r_{+++}/(2p)]^2}dx\;x^\eta \left(A_{++}-B_{++} \right) \right. \nonumber\\
    &\left. \qquad - (-1)^{h+u}(r_2-r_3)^{t+v}\int_{[r_{--+}/(2p)]^2}^{[r_{+-+}/(2p)]^2}dx\;x^\eta \left(A_{-+}-B_{-+} \right) \right. \nonumber\\
    &\left. \qquad - (-r_2+r_3)^{t+v}\int_{[r_{-+-}/(2p)]^2}^{[r_{++-}/(2p)]^2}dx\;x^\eta \left(A_{+-}-B_{+-} \right) \right. \nonumber \\
    &\left. \qquad + (r_2+r_3)^{t+v}\int_{[r_{---}/(2p)]^2}^{[r_{+--}/(2p)]^2}dx\;x^\eta \left(A_{--}-B_{--} \right) \right],
\end{align}
where we have defined $\eta \equiv [g+2m-t-u-v-1]/2$ and 
\begin{align}
    r_{\pm \pm \pm} \equiv \pm r_1 \pm r_2 \pm r_3.
\end{align}

By evaluating the integrals of equation (\ref{eqn: IA_1sub}), we obtain a final result for $I_{\text{pl}}$:\footnote{https://functions.wolfram.com/GammaBetaErf/Gamma2/21/01/02/01/0001/}
\begin{align}
\label{eqn:IA_final}
    I_{\text{pl}}(r_1,r_2,r_3;p) &= \Gamma\left(\frac{\alpha+1}{2}\right)\frac{ip^3}{r_1r_2r_3}(2ip)^\zeta \sum\limits_{b=0}^{\ell_3}\binom{\ell_3}{b}(-1)^br_3^{-2b}\sum\limits_{c=0}^{2b}\binom{2b}{c}(2ip)^{2b-c}(-1)^{c-\beta}\sum\limits_{d=0}^{\ell_2}\binom{\ell_2}{d} \nonumber \\
    & \qquad \times (-1)^dr_2^{-2d}\sum\limits_{g=0}^{2d}\binom{2d}{g}(2p)^{2d-g}\sum\limits_{h=0}^c\binom{c}{h}(2p)^{c-h}r_3^h\frac{1}{\epsilon+1}\sum\limits_{m=0}^{\ell_1}\binom{\ell_1}{m}(-1)^mr_1^{-2m} \nonumber \\
    & \qquad \times \sum\limits_{t=0}^{2m}\binom{2m}{t}(2p)^{2m-t}\sum\limits_{u=0}^{g}\binom{g}{u}(-1)^{g-u}r_3^u \sum\limits_{v=0}^{g-u}\binom{g-u}{v}(2p)^{g-u-v} \nonumber \\
    & \qquad \times \Big[(-1)^{h+u}(-r_2-r_3)^{t+v}\left(Z_{+++}-Z_{-++}-Y_{+++}+Y_{-++} \right) \Big. \nonumber\\
    &\Big. \qquad - (-1)^{h+u}(r_2-r_3)^{t+v}\left(Z_{+-+}-Z_{--+}-Y_{+-+}+Y_{--+} \right) \Big. \nonumber\\
    &\Big. \qquad - (-r_2+r_3)^{t+v}\left(Z_{++-}-Z_{-+-}-Y_{++-}+Y_{-+-} \right) \Big. \nonumber \\
    &\Big. \qquad + (r_2+r_3)^{t+v}\left(Z_{+--}-Z_{---}-Y_{+--}+Y_{---} \right) \Big],
\end{align}
where we have defined
\begin{align}
    Z_{\pm\pm\pm} &\equiv \frac{1}{\eta+\epsilon+2}\left[\left(\frac{r_{\pm\pm\pm}}{2p}\right)^{2(\eta+\epsilon+2)}\Gamma\left(\beta+1,\left(\frac{r_{\pm\pm\pm}}{2p}\right)^2 \right) \right. \nonumber \\
    &\left. \qquad\qquad\qquad -\Gamma\left(\beta+\eta+\epsilon+3,\left(\frac{r_{\pm\pm\pm}}{2p}\right)^2 \right) \right]
\end{align}
and
\begin{align}
    Y_{\pm\pm\pm} &\equiv \frac{1}{\eta+1}\left[\left(\frac{r_{\pm\pm\pm}}{2p}\right)^{2(\eta+1)}\Gamma\left(\beta+\epsilon+2,\left(\frac{r_{\pm\pm\pm}}{2p}\right)^2 \right) \right. \nonumber \\
    &\left. \qquad\qquad - \Gamma\left(\beta+\eta+\epsilon+3,\left(\frac{r_{\pm\pm\pm}}{2p}\right)^2 \right) \right].
\end{align}
We have suppressed the arguments of $Z_{\pm\pm\pm}$ and $Y_{\pm\pm\pm}$.

We now turn to $I_{\text{hg}}$, given in equation (\ref{eqn:IB_orig}) and duplicated below:
\begin{align}
\label{eqn:IB_orig_dup}
    I_{\text{hg}}(r_1,r_2,r_3;p) &= \frac{2}{\alpha+1}(-1)^{(\alpha+1)/2}(2p)^{-(\alpha+1)}\int_{-1}^1dq_1\;(1-q_1^2)^{\ell_1}\int_{-1}^1dq_2\;(1-q_2^2)^{\ell_2}\nonumber \\ &\qquad\qquad\qquad\qquad \times \int_{-1}^1dq_3\;(1-q_3^2)^{\ell_3}s_{123}^{n+L+1}\;_1{F_1}\left(1;\frac{\alpha+3}{2}; -\left(\frac{s_{123}}{2p} \right)^2 \right).
\end{align}
\\
\indent To evaluate the innermost, $q_3$ integral, we must recall that $s_{123}$ depends on $q_3$ (equation \ref{eqn:s123}). We make the change of variable $x = -[s_{123}/(2p)]^2$, which is the argument of the hypergeometric function in equation (\ref{eqn:IB_orig_dup}). Recalling the definition of $s_{12}$ from equation (\ref{eqn:s12}), we use $q_3 = [2pi\sqrt{x}-s_{12}]/r_3$ to rewrite the $(1-q_3^2)^{\ell_3}$ factor in equation (\ref{eqn:IB_orig_dup}) before using a binomial expansion on it:
\begin{align}
\label{eqn:IB_3sub}
    I_{\text{hg}}(r_1,r_2,r_3;p) &= \frac{-1}{(\alpha+1)r_3}(-1)^{(\alpha+1)/2}i^{n+L}(2p)^{\zeta+1}\sum\limits_{b=0}^{\ell_3}\binom{\ell_3}{b}(-1)^br_3^{-2b}\sum\limits_{c=0}^{2b}\binom{2b}{c}(2ip)^{2b-c}(-1)^c \nonumber\\
    & \qquad \times \int_{-1}^1dq_1\;(1-q_1^2)^{\ell_1}\int_{-1}^1dq_2\;(1-q_2^2)^{\ell_2}s_{12}^c\int_{-[(s_{12}-r_3)/(2p)]^2}^{-[(s_{12}+r_3)/(2p)]^2}dx\;x^\kappa \nonumber \\
    & \qquad\qquad\qquad\qquad\qquad\qquad\qquad\qquad\qquad\qquad \times \;_1{F_1}\left(1;\frac{\alpha+3}{2};x\right),
\end{align}
where we have defined $\kappa \equiv [n+L+2b-c]/2$. We now integrate over $x$, resulting in\footnote{https://functions.wolfram.com/HypergeometricFunctions/Hypergeometric1F1/21/01/02/01/0002/}
\begin{align}
\label{eqn:IB_3}
    I_{\text{hg}}(r_1,r_2,r_3;p) &= \frac{-1}{(\alpha+1)r_3}(-1)^{(\alpha+1)/2}i^{n+L}(2p)^{\zeta+1}\sum\limits_{b=0}^{\ell_3}\binom{\ell_3}{b}(-1)^br_3^{-2b}\sum\limits_{c=0}^{2b}\binom{2b}{c}(2ip)^{2b-c}(-1)^c \nonumber \\
    & \qquad \times \frac{1}{\kappa+1}(-1)^{\kappa+1}\int_{-1}^1dq_1\;(1-q_1^2)^{\ell_1}\int_{-1}^1dq_2\;(1-q_2^2)^{\ell_2}s_{12}^c \nonumber \\
    & \qquad \times \left[\left(\frac{s_{12}+r_3}{2p}\right)^{2(\kappa+1)}\;_2{F_2}\left(1,\kappa+1;\frac{\alpha+3}{2},\kappa+2;-\left(\frac{s_{12}+r_3}{2p}\right)^2 \right) \right. \nonumber \\
    & \left. \qquad - \left(\frac{s_{12}-r_3}{2p}\right)^{2(\kappa+1)}\;_2{F_2}\left(1,\kappa+1;\frac{\alpha+3}{2},\kappa+2;-\left(\frac{s_{12}-r_3}{2p}\right)^2 \right) \right].
\end{align}

\indent To evaluate the innermost, $q_2$ integral of equation (\ref{eqn:IB_3}), we use the change of variable \\$x = -[(s_{12} \pm r_3)/(2p)]^2$, where the $\pm$ sign represents the sign of $r_3$ in the arguments of the hypergeometric functions in equation (\ref{eqn:IB_3}). Since $q_2 = [2ip\sqrt{x}-r_1q_1 \mp r_3]/r_2$, we can rewrite the $s_{12}^c$ factor in equation (\ref{eqn:IB_3}) as $s_{12}^c = (2ip\sqrt{x} \mp r_3)^c$, then use a binomial expansion on it. We also use a binomial expansion on the $(1-q_2^2)^{\ell_2}$ factor of equation (\ref{eqn:IB_3}) after writing it in terms of $x$:
\begin{align}
\label{eqn:IB_2sub}
    I_{\text{hg}}(r_1,r_2,r_3;p) &= \frac{-ip}{(\alpha+1)r_2r_3}(-1)^{(\alpha+1)/2}i^{n+L}(2p)^{\zeta+1}\sum\limits_{b=0}^{\ell_3}\binom{\ell_3}{b}(-1)^br_3^{-2b}\sum\limits_{c=0}^{2b}\binom{2b}{c}(2ip)^{2b-c}(-1)^c \nonumber \\
    & \qquad \times \frac{1}{\kappa+1}\sum\limits_{d=0}^{\ell_2}\binom{\ell_2}{d}(-1)^dr_2^{-2d}\sum\limits_{g=0}^{2d}\binom{2d}{g}(2ip)^{2d-g}\sum\limits_{h=0}^c\binom{c}{h}(2ip)^{c-h}r_3^h \nonumber \\
    & \qquad \times \int_{-1}^1dq_1\;(1-q_1^2)^{\ell_1}\left[(-1)^h(-r_1q_1-r_3)^g\int_{-[r_{q-+}/(2p)]^2}^{-[r_{q++}/(2p)]^2}dx\;x^{\lambda} \right. \nonumber \\
    &\left. \qquad\qquad\qquad\qquad\qquad\qquad\qquad\qquad\qquad\qquad\qquad \times _2{F_2}\left(1,\kappa+1;\frac{\alpha+3}{2},\kappa+2;x \right) \right. \nonumber \\
    &\left. \qquad -(-r_1q_1+r_3)^g\int_{-[r_{q--}/(2p)]^2}^{-[r_{q+-}/(2p)]^2}dx\;x^{\lambda}\;_2{F_2}\left(1,\kappa+1;\frac{\alpha+3}{2},\kappa+2;x \right) \right],
\end{align}
where we recall $r_{q \pm \pm} \equiv r_1q_1 \pm r_2 \pm r_3$ and define $\lambda \equiv \kappa+1+[c+2d-g-h-1]/2$. After evaluating the integrals of equation (\ref{eqn:IB_2sub}) with respect to $x$, we obtain\footnote{https://functions.wolfram.com/HypergeometricFunctions/HypergeometricPFQ/21/01/02/01/0001/}
\begin{align}
\label{eqn:IB_2}
    I_{\text{hg}}(r_1,r_2,r_3;p) &= \frac{-ip}{(\alpha+1)r_2r_3}(-1)^{(\alpha+1)/2}i^{n+L}(2p)^{\zeta+1}\sum\limits_{b=0}^{\ell_3}\binom{\ell_3}{b}(-1)^br_3^{-2b}\sum\limits_{c=0}^{2b}\binom{2b}{c}(2ip)^{2b-c}(-1)^c \nonumber \\
    & \qquad \times \frac{1}{\kappa+1}\sum\limits_{d=0}^{\ell_2}\binom{\ell_2}{d}(-1)^dr_2^{-2d}\sum\limits_{g=0}^{2d}\binom{2d}{g}(2ip)^{2d-g}\sum\limits_{h=0}^c\binom{c}{h}(2ip)^{c-h}r_3^h\frac{1}{\lambda+1} \nonumber \\
    & \qquad \times (-1)^{\lambda+1}\int_{-1}^1dq_1\;(1-q_1^2)^{\ell_1} \Big[(-1)^h(-r_1q_1-r_3)^g\left(C_{++}-C_{-+}\right) \Big. \nonumber \\
    &\Big. \qquad -(-r_1q_1+r_3)^g\left(C_{+-}-C_{--}\right) \Big],
\end{align}
where we have defined (with arguments suppressed)
\begin{align}
    C_{\pm\pm} \equiv \left(\frac{r_{q \pm \pm}}{2p} \right)^{2(\lambda+1)}\;_3{F_3}\left(\lambda+1,1,\kappa+1;\lambda+2,\frac{\alpha+3}{2},\kappa+2;-\left(\frac{r_{q \pm \pm}}{2p}\right)^2 \right).
\end{align}

\indent To evaluate the $q_1$ integral, we use the change of variable $x = -[r_{q\pm\pm}/(2p)]^2$. We use \\ $q_1 = [2ip\sqrt{x} \mp r_2 \mp r_3]/r_1$ to rewrite the $(1-q_1^2)^{\ell_1}$ factor in equation (\ref{eqn:IB_2}), then use a binomial expansion on it. We also rewrite the $(-r_1q_1 \mp r_3)^g$ factor of equation (\ref{eqn:IB_2}) in terms of $x$, then use another binomial expansion on this factor:
\begin{align}
\label{eqn:IB_1sub}
    I_{\text{hg}}(r_1,r_2,r_3;p) &= \frac{p^2}{(\alpha+1)r_1r_2r_3}(-1)^{(\alpha+1)/2}i^{n+L}(2p)^{\zeta+1}\sum\limits_{b=0}^{\ell_3}\binom{\ell_3}{b}(-1)^br_3^{-2b}\sum\limits_{c=0}^{2b}\binom{2b}{c}(2ip)^{2b-c} \nonumber \\
    & \qquad \times (-1)^c\frac{1}{\kappa+1}\sum\limits_{d=0}^{\ell_2}\binom{\ell_2}{d}(-1)^dr_2^{-2d}\sum\limits_{g=0}^{2d}\binom{2d}{g}(2ip)^{2d-g}\sum\limits_{h=0}^c\binom{c}{h}(2ip)^{c-h}r_3^h\frac{1}{\lambda+1} \nonumber \\
    & \qquad \times (-1)^{\lambda+1}\sum\limits_{m=0}^{\ell_1}\binom{\ell_1}{m}(-1)^mr_1^{-2m}\sum\limits_{t=0}^{2m}\binom{2m}{t}(2ip)^{2m-t}\sum\limits_{u=0}^g\binom{g}{u}r_3^u(-1)^{g-u} \nonumber \\
    & \qquad \times \sum\limits_{v=0}^{g-u}\binom{g-u}{v}(2ip)^{g-u-v}\left[(-1)^{h+u}(-r_2-r_3)^{t+v}\int_{-[r_{-++}/(2p)]^2}^{-[r_{+++}/(2p)]^2}dx\;x^{\phi}C_{++} \right. \nonumber\\
    &\left. \qquad -(-1)^{h+u}(r_2-r_3)^{t+v}\int_{-[r_{--+}/(2p)]^2}^{-[r_{+-+}/(2p)]^2}dx\;x^{\phi}C_{-+} \right. \nonumber\\
    &\left. \qquad -(-r_2+r_3)^{t+v}\int_{-[r_{-+-}/(2p)]^2}^{-[r_{++-}/(2p)]^2}dx\;x^{\phi}C_{+-} \right. \nonumber \\
    &\left. \qquad + (r_2+r_3)^{t+v}\int_{-[r_{---}/(2p)]^2}^{-[r_{+--}/(2p)]^2}dx\;x^{\phi}C_{--} \right],
\end{align}
where we recall $r_{\pm\pm\pm} \equiv \pm r_1 \pm r_2 \pm r_3$ and define $\phi \equiv [g+2m-t-u-v-1]/2$. After evaluating the integrals in equation (\ref{eqn:IB_1sub}), we obtain a final result for $I_{\text{hg}}$:\footnote{https://functions.wolfram.com/HypergeometricFunctions/HypergeometricPFQ/21/01/02/01/0001/}
\begin{align}
\label{eqn:IB_final}
    I_{\text{hg}}(r_1,r_2,r_3;p) &= \frac{p^2}{(\alpha+1)r_1r_2r_3}(-1)^{(\alpha+1)/2}i^{n+L}(2p)^{\zeta+1}\sum\limits_{b=0}^{\ell_3}\binom{\ell_3}{b}(-1)^br_3^{-2b}\sum\limits_{c=0}^{2b}\binom{2b}{c}(2ip)^{2b-c} \nonumber \\
    & \qquad \times (-1)^c\frac{1}{\kappa+1}\sum\limits_{d=0}^{\ell_2}\binom{\ell_2}{d}(-1)^dr_2^{-2d}\sum\limits_{g=0}^{2d}\binom{2d}{g}(2ip)^{2d-g}\sum\limits_{h=0}^c\binom{c}{h}(2ip)^{c-h}r_3^h\frac{1}{\lambda+1} \nonumber \\
    & \qquad \times \sum\limits_{m=0}^{\ell_1}\binom{\ell_1}{m}(-1)^mr_1^{-2m}\sum\limits_{t=0}^{2m}\binom{2m}{t}(2ip)^{2m-t}\sum\limits_{u=0}^g\binom{g}{u}r_3^u(-1)^{g-u}\sum\limits_{v=0}^{g-u}\binom{g-u}{v} \nonumber \\
    & \qquad \times (2ip)^{g-u-v}\frac{1}{\rho+1}(-1)^{\rho+1}\Big[(-1)^{h+u}(-r_2-r_3)^{t+v}\left(F_{+++}-F_{-++}\right) \Big. \nonumber\\
    &\Big. \qquad -(-1)^{h+u}(r_2-r_3)^{t+v}\left(F_{+-+}-F_{--+}\right) - (-r_2+r_3)^{t+v}\left(F_{++-}-F_{-+-}\right) \Big. \nonumber \\
    &\Big. \qquad +(r_2+r_3)^{t+v}\left(F_{+--}-F_{---}\right) \Big],
\end{align}
where we have defined $\rho \equiv \lambda+1+[g+2m-t-u-v-1]/2$ and 
\begin{align}
    F_{\pm\pm\pm} \equiv \left(\frac{r_{\pm\pm\pm}}{2p}\right)^{2(\rho+1)}\;_4{F_4}\left(\rho+1,\lambda+1,1,\kappa+1;\rho+2,\lambda+2,\frac{\alpha+3}{2},\kappa+2;-\left(\frac{r_{\pm\pm\pm}}{2p}\right)^2 \right).
\end{align}
The arguments of $F_{\pm\pm\pm}$ have been suppressed.

\indent The integral of three sBFs with a Gaussian was given in terms of $I_{\text{pl}}$ and $I_{\text{hg}}$ in equation (\ref{eqn:f_in_terms_of_AB}); we duplicate this below: 
\begin{align}
\label{eqn:f_as_IA_IB_dup}
    f_{\ell_1,\ell_2,\ell_3}(r_1,r_2,r_3;p) &= \frac{2^{-L-4}}{\Gamma(\ell_1+1)\Gamma(\ell_2+1)\Gamma(\ell_3+1)}\;r_1^{\ell_1}r_2^{\ell_2}r_3^{\ell_3}\sum\limits_{\zeta=0}^{n+L}\binom{n+L}{\zeta}\left(\frac{i}{2p^2}\right)^\zeta p^{-(\alpha+1)} \nonumber \\
    & \qquad \times (I_{\text{pl}}-I_{\text{hg}}).
\end{align}
We now use the results for $I_{\text{pl}}$ (equation \ref{eqn:IA_final}) and $I_{\text{hg}}$ (equation \ref{eqn:IB_final}) to evaluate equation (\ref{eqn:f_as_IA_IB_dup}):
{\color{purple}
\begin{align}
\label{eqn:3SBF_GAUSS_FINAL}
    f_{\ell_1,\ell_2,\ell_3}(r_1,r_2,r_3;p) &= \frac{2^{-L-4}p^3}{\Gamma(\ell_1+1)\Gamma(\ell_2+1)\Gamma(\ell_3+1)}r_1^{\ell_1-1}r_2^{\ell_2-1}r_3^{\ell_3-1}\sum\limits_{\zeta=0}^{n+L}\binom{n+L}{\zeta}i^\zeta p^{-(n+L+1)} \nonumber \\
    & \qquad \times \sum\limits_{b=0}^{\ell_3}\binom{\ell_3}{b}r_3^{-2b}\sum\limits_{c=0}^{2b}\binom{2b}{c}i^{2b-c}\sum\limits_{d=0}^{\ell_2}\binom{\ell_2}{d}r_2^{-2d}\sum\limits_{g=0}^{2d}\binom{2d}{g}\sum\limits_{h=0}^c\binom{c}{h}r_3^h \nonumber \\
    & \qquad \times \sum\limits_{m=0}^{\ell_1}\binom{\ell_1}{m}r_1^{-2m}\sum\limits_{t=0}^{2m}\binom{2m}{t}\sum\limits_{u=0}^g\binom{g}{u}r_3^u(-1)^{\Omega}\sum\limits_{v=0}^{g-u}\binom{g-u}{v}(2p)^{\sigma} \Big\{(-1)^{h+u} \nonumber \\
    & \qquad \times (-r_2-r_3)^{t+v}\Big[\omega\left(Z_{+++}-Z_{-++} -Y_{+++}+Y_{-++}\right) -\psi\left(F_{+++} \right.\Big. \nonumber \\
    & \qquad \left.\Big. -F_{-++}\right)\Big] - (-1)^{h+u}(r_2-r_3)^{t+v}\Big[\omega\left(Z_{+-+} -Z_{--+}-Y_{+-+}+Y_{--+} \right) \Big.\Big. \nonumber \\
    &\Big.\Big. \qquad -\psi\left(F_{+-+}-F_{--+}\right) \Big]  -(-r_2+r_3)^{t+v}\Big[\omega\left(Z_{++-}-Z_{-+-}-Y_{++-} \right.\Big. \nonumber \\
    &\left. \Big. \qquad +Y_{-+-}\right) -\psi\left(F_{++-}-F_{-+-}\right)\Big] +(r_2+r_3)^{t+v}\Big[\omega\left(Z_{+--}-Z_{---} \right.\Big. \nonumber \\
    & \left.\Big. \qquad -Y_{+--}+Y_{---}\right) -\psi\left(F_{+--}-F_{---}\right) \Big] \Big\}\textcolor{black}{,}
\end{align}}
where we have defined
\begin{align}
    \Omega &\equiv b+c+d+g+m-u, \nonumber \\
    \sigma &\equiv 2b+2d+2m-h-t-u-v, \nonumber \\
    \omega &\equiv \Gamma\left(\frac{\alpha+1}{2}\right)\frac{(-1)^{-\beta} i^{\zeta+1}}{\epsilon+1}, \nonumber \\
    \tau &\equiv n+L+c+2d+2m-h-t-u-v, \nonumber \\
    \psi &\equiv \frac{2i^\tau}{(\alpha+1)(\kappa+1)(\lambda+1)(\rho+1)}(-1)^{\rho+1+(\alpha+1)/2}.
\end{align}

Equation (\ref{eqn:3SBF_GAUSS_FINAL}) is our final result for the triple-sBF integral with Gaussian damping weighted by any non-negative integer power of $k$, given by equation (\ref{eqn:bowman_orig}). This final result is in terms of incomplete Gamma functions and hypergeometric functions; the base cases of the recursion relation for the triple-sBF integral with Gaussian damping in \S \ref{sec:gauss_rec} also depend on the same types of functions. Additionally, both these base cases and equation (\ref{eqn:3SBF_GAUSS_FINAL}) have a similar structure in front of the incomplete Gamma and hypergeometric functions: powers of $(\mp r_2 \mp r_3)$. 

The recursion relation in \S \ref{sec:gauss_rec} was limited to triple-sBF integrals with Gaussian damping and the Jacobian $k^2dk$. In \S \ref{sec:para_diff}, it was shown that parametric differentiation can be used to extend this recursion to integrals weighted by $k$ to any even power $\geq 2$. The result given in equation (\ref{eqn:3SBF_GAUSS_FINAL}) allows more freedom in the power of $k$; with this result, we are able to evaluate triple-sBF integrals with Gaussian damping weighted by 
$k$ to any non-negative integer power $\geq 0$.

The steps outlined in this section may be used to evaluate integrals of any number of sBFs with Gaussian damping. The sBFs can be rewritten as integrals over $q_i$, as we have done using equation (\ref{eqn:hankel_contour}). Then, as in equations (\ref{eqn:bowman_Ik}) and (\ref{eqn:bowman_Ik_final}), the Gaussian integral with respect to $k$ can be evaluated. Finally, the remaining $q_i$ integrals will require many changes of variables and binomial expansions, following the work in this section to give a result in terms of incomplete Gamma and hypergeometric functions.

\section{Computational Complexity}
\subsection{Our Methods Compared to Direct Integration}
\label{ssec:comp_complexity}
Our analytic results for triple-sBF integrals with exponential (base cases of \S \ref{sssec:exp_rec}) and Gaussian damping (base cases of \S \ref{sec:gauss_rec} as well as equation \ref{eqn:3SBF_GAUSS_FINAL}) can be computed more efficiently than a direct integration of damped triple-sBF integrals. For a direct integration, one would need a vector of $N_k$ $k$ points to perform the integral at each triple $\{r_1,r_2,r_3\}$ on a three-dimensional grid of the latter. If there are $N_r$ sample points for each $r_i$, the total cost of a direct integration is $N_k N_r^3$. We now turn to the computational complexity of our methods.

The Legendre functions of the second kind in the analytic result for the triple-sBF integral with exponential damping (equations \ref{eqn:exp_bc1}, \ref{eqn:exp_bc2}, and \ref{eqn:exp_bc3}) all depend on $R_{\pm \pm} \equiv \left(-ip^2 \pm r_2 \pm r_3 \right)/r_1$ (equation \ref{eqn:R_def}). Given values for $r_1$, $r_2$, and $r_3$ (the free arguments of the sBFs), a new variable $\mathcal{R}$ can be defined such that it ranges from the minimum of $R_{\pm \pm}$ to the maximum of $R_{\pm \pm}$. We may then map all required $R_{\pm \pm}$ to $\mathcal{R}$. Once this is done, the Legendre functions of the second kind may be evaluated on a one-dimensional vector of $\mathcal{R}$. The numerator of $R_{\pm\pm}$ causes this $\mathcal{R}$ grid to be be $4N_r$ in length. However, the $r_1$ in the denominator will shrink this grid, since in a typical covariance calculation, the minimum $r_1$ is larger than 1 Mpc and so dividing by it will reduce the range covered by the numerator. The length of the $\mathcal{R}$ grid stems from our method's results rotating $r_1$, $r_2$, and $r_3$ to $45^{\circ}$ lines coming out of the origin of that coordinate system. The integrals only depend on the values along these lines. This concept was previously used in a numerical algorithm for sBF integrals (\cite{rotation}, Figure 1). The computational cost of our method thus scales as $N_r$ rather than $N_kN_r^3$. 

The incomplete Gamma and hypergeometric functions in the results for the triple-sBF integral with Gaussian damping (equations \ref{eqn:gauss_bc1}, \ref{eqn:gauss_bc2}, \ref{eqn:gauss_bc3}, and \ref{eqn:3SBF_GAUSS_FINAL}) all depend on $r_{\pm \pm \pm } \equiv \pm r_1 \pm r_2 \pm r_3$. Similarly to $\mathcal{R}$, a new variable $\mathfrak{r}$ ranging from the minimum of $r_{\pm \pm \pm }$ to the maximum of $r_{\pm \pm \pm }$ can be defined. The incomplete gamma and hypergeometric functions can be evaluated on a one-dimensional $\mathfrak{r}$ vector with length $6N_r$. As the computational cost of our method scales as $N_r$, it will be orders of magnitude faster than a direct numerical integration, even when accounting for the prefactors that arise in our method due to summing over a number of incomplete gamma and hypergeometric functions.

Additionally, the triple-sBF integrals do not need to be calculated explicitly for every $\{\ell_1,\ell_2,\ell_3\}$ and $\{r_1,r_2,r_3\}$ combination. The recursion allows triple-sBF integrals to be expressed in terms of three base cases. If any of these base cases have already been evaluated and the result stored, less computations are required to obtain the desired integral. If, for example, all three required base cases have previously been evaluated, the recursion method is simply a combination of the base cases rather than a method with computational complexity $N_r$. Thus, the computational cost of the recursion method will scale at most as $N_r$, but will often be less costly.

\subsection{Comparison of Results for the Gaussian-Damped triple-sBF Integral}
We have derived results for the Gaussian-damped triple sBF integral in two different ways: with a recursion relation (\S \ref{sec:gauss_rec}) and by the Hankel-Bowman method (\S \ref{sec:bowman}). The recursion relation can be used to evaluate Gaussian-damped triple-sBF integrals with the Jacobian $k^2dk$ (\S \ref{sec:gauss_rec}). By parametric differentiation (\S \ref{sec:para_diff}), this recursion method can be extended to integrals involving $k$ to any even power $\geq 2$. The Hankel-Bowman method (\S \ref{sec:bowman}), however, can be used to evaluate Gaussian-damped triple-sBF integrals with the Jacobian $k^ndk$, as long as $n$ is a non-negative integer. Thus, a major advantage of the Hankel-Bowman method is that it can be used to evaluate a wider range of integrals. 

The recursion method requires at least three base cases (equations \ref{eqn:gauss_bc1}, \ref{eqn:gauss_bc2}, and \ref{eqn:gauss_bc3}), each of which depend on eight incomplete gamma functions and eight regularized hypergeometric functions. As demonstrated in \S \ref{sec:stability}, using the recursion to evaluate integrals with higher-order sBFs will require more than three base cases; therefore, each recursion will require computation of at least 24 incomplete gamma functions and at least 24 regularized hypergeometric functions. 

The result for the Hankel-Bowman method (equation \ref{eqn:3SBF_GAUSS_FINAL}) appears to depend upon 32 incomplete gamma functions and eight hypergeometric functions. However, the parameters of these functions depend on $\alpha$ (ranging from $0$ to $n+\ell_1+\ell_2+\ell_3$), as well as the values of the variables being summed over in equation (\ref{eqn:3SBF_GAUSS_FINAL}). Thus, in most cases, more than 32 incomplete gamma functions and more than eight hypergeometric functions will need to be computed for the Hankel-Bowman method. The recursion method will therefore typically be more efficient than the Hankel-Bowman method for evaluating Gaussian-damped triple-sBF integrals.

\section{Conclusion}
We have shown that the spherical Bessel function recursion relation can be used to obtain a recursion for triple-sBF integrals with exponential or Gaussian damping, generalizing an approach pioneered by \cite{wk}. This approach requires three base cases to anchor it; once those base cases are determined, any triple-sBF integral with damping can be evaluated as long as the (integer) orders of the sBFs are at least zero, the argument of one of the sBFs is positive, and the arguments of the other two sBFs are nonzero. The orders of the sBFs and their arguments must also be real. The recursion allows the triple-sBF integral with exponential damping to be written in a simple form dependent upon Legendre functions of the second kind. For three sBFs with Gaussian damping, the recursion shows that the integral can be evaluated in terms of regularized hypergeometric and incomplete Gamma functions. 

For the case with exponential damping, the base cases depend on Legendre functions with complex arguments. Each base case requires four Legendre functions (two negative complex conjugate pairs). By expanding the Legendre functions in an infinite series, we demonstrated that the base cases are indeed real, as expected given the reality of the integrand and of the domain of integration.

The work shown here is new; previously, a recursion relation had only been used to evaluate triple-sBF integrals with no damping term \cite{wk}; this is the $p=0$ limit of our work (which is the same for the exponentially-damped and Gaussian cases). Notably, the previous work also required that the free arguments of the three sBFs, $r_i$, be able to form a closed triangle, and also that the sum of the three sBF orders, $\ell_i$, be even. The results of the present work require only that $\{p, \ell_i,r_i\} \in \mathbb{R}$, $\ell_i \in \mathbb{Z}$, $\ell_i \geq 0$, $r_1 > 0$, and $r_i \neq 0$, but demand no constraints between the individual $r_i$ and $\ell_i$.

The stability of the recursion depends on two factors: whether the base cases can be evaluated accurately and whether as many base cases as needed can be co-added without loss of numerical precision. Accurate numerical algorithms can be used to evaluate the special functions within the base cases. Numerical precision can be handled with existing libraries or balanced sum algorithms, which are useful for adding values with order of magnitude differences. The recursions are thus stable for both the exponentially- and Gaussian-damped integrals.

Additionally, the recursion relations for triple-sBF integrals with the Jacobian $k^2dk$ and exponential or Gaussian damping can be used to obtain results for such integrals weighted by higher powers of $k$. The partial derivative with respect to $-p^2$ of the damped triple-sBF integral can be moved inside the integral so that it acts on the damping. Each partial derivative increases the power of k by one for exponential damping and by two for Gaussian damping. Thus, by first evaluating the damped triple-sBF integral using the recursion relations outlined in \S \ref{sssec:exp_rec} and \S \ref{sec:gauss_rec} and then taking the necessary amount of partial derivatives with respect to $-p^2$, the damped triple-sBF integrals can be evaluated for $k$ to any integer power $\geq 2$ for exponential damping and $k$ to any even power $\geq 2$ for Gaussian damping.

Finally, we have found a new method (the Hankel-Bowman method) to evaluate triple-sBF integrals with Gaussian damping weighted by any non-negative integer power of $k$. This method requires that $\{p,\ell_i,r_i\} \in \mathbb{R}$, $\{\ell_i,n\} \in \mathbb{Z}$, $\ell_i \geq 0$, $r_i\neq 0$, $p \neq 0$, and $n \geq 0$. The sBFs are rewritten as integrals derived from Hankel's contour integral, then many changes of variable and binomial expansions are used to obtain a result in terms of incomplete Gamma functions and hypergeometric functions. This method can also be used to evaluate integrals of any number of sBFs with Gaussian damping. While the recursion relation for the triple-sBF integral with Gaussian damping only can be used for the Jacobian $k^2dk$, and extended by parametric differentiation to even powers of $k$ that are greater than or equal to two, the work outlined in \S \ref{sec:bowman} allows such integrals weighted by any non-negative integer power of $k$ to be evaluated.

We have demonstrated that our methods for Gaussian-damped triple-sBF integrals are more computationally efficient than a direct integration: our methods scale linearly as $N_r$, while a direct integration would scale as $N_kN_r^3$. The recursion method for Gaussian-damped triple-sBF integrals (\S \ref{sec:gauss_rec}) requires computation of a minimum of 24 incomplete Gamma functions and 24 hypergeometric functions. The Hankel-Bowman method (\S \ref{sec:bowman}) requires at least 32 incomplete Gamma functions and eight hypergeometric functions to be evaluated. However, due to the many finite sums in the final result of the Hankel-Bowman method (equation \ref{eqn:3SBF_GAUSS_FINAL}), typically many more incomplete Gamma and hypergeometric functions will need to be computed. Therefore, it is preferable to use the recursion method over the Hankel-Bowman method for Gaussian-damped triple-sBF integrals with the Jacobian $k^ndk$, where $n$ is an even power $\geq 2$.  

\section*{Acknowledgments}
We acknowledge useful conversations with the Slepian research group and especially thank Bob Cahn for many fruitful discussions on sBF integrals over a long period. 

\bibliographystyle{unsrt}
\bibliography{refs}

\end{document}